\documentclass[manuscript]{aastex}
\usepackage{emulateapj5}
\usepackage{apjfonts}

\newcommand{\kband}{$K$-band}

\newcommand{\kms}{km s$^{-1}$}

\slugcomment{to appear in the Astrophysical Journal}

\shorttitle{PDS 456: Luminous Infrared QSO}
\shortauthors{Yun et al.}

\begin{document}


\title{Multi-wavelength Observations of the Gas-rich Host Galaxy of PDS~456: 
a New Challenge for the ULIRG-to-QSO Transition Scenario}

\author{Min S. Yun}
\affil{Astronomy Department, University of Massachusetts,
    Amherst, MA 01003, USA}
\email{myun@astro.umass.edu}

\author{N. A. Reddy, N. Z. Scoville}
\affil{California Institute of Technology, Pasadena, CA 91125, USA}
\email{nar@astro.caltech.edu, nzs@astro.caltech.edu}

\author{D. T. Frayer}
\affil{SIRTF Science Center, California Institute of Technology, 
220--06, Pasadena, CA 91125, USA}
\email{frayer@ipac.caltech.edu}

\and

\author{E. I. Robson\altaffilmark{1}, R. P. J. Tilanus}
\affil{Joint Astronomy Centre, 660 N. Aohoku Place, Hilo, HI 96720, USA}
\email{eir@roe.ac.uk, r.tilanus@jach.hawaii.edu}

\altaffiltext{1}{Current address: Royal Observatory, Blackford Hill, Edinburgh
EH9 3HJ}

\begin{abstract}
We report new \kband, radio continuum, and 
CO (1--0) imaging observations and 850 $\mu$m photometric observations
of PDS~456, the most luminous QSO in the local universe ($z<0.3$).
The $0.6''$ resolution \kband\ image obtained using the Keck telescope
shows three compact $m_K\sim 16.5$ ($M_K\sim -21$) sources  
at a projected distance of $\sim 10$ kpc to the southwest, and 
the host galaxy of PDS~456 may be interacting or merging
with one or more companions.  The observations using
the OVRO millimeter array has revealed a narrow CO (1--0) line 
(FWHM = 181 \kms) centered at $z=0.1849$, and $9\times 10^9 M_\odot$ 
of molecular gas mass is inferred.  Radio continuum luminosity is 
nearly an order of magnitude larger than expected from its FIR
luminosity, and the radio source, unresolved
by the 2\arcsec\ beam of the VLA, is dominated by the AGN
activity.  Our 850 $\mu$m photometric observations 
suggest that the cold dust content of the host galaxy is less than 
one half of the amount in Arp~220. 

The analysis of the spectral energy 
distribution  reveals both a QSO-like and a ULIRG-like nature, 
and the observed IR, X-ray, and gas properties suggest that the AGN
activity dominates its luminosity.  
PDS~456 displays many characteristics expected of an object
undergoing a transition from an ultra-luminous infrared galaxy (ULIRG)
to a classical QSO phase as proposed by Sanders et al., 
including an optical spectrum dominated by broad emission lines, 
large X-ray and IR luminosity, a large cold gas/dust content, and an
extremely large $L_{FIR}/M_{H_2}$ ratio ($\gtrsim 100~L_\odot/M_\odot$). 
ULIRGs and IR QSOs form a broad continuous track in the ``star 
formation efficiency'' plot in the manner consistent with the
ULIRG-QSO transition scenario, relating the evolution in the dust 
processed luminosity with the available fuel (gas and dust) supply.
However, the location of PDS~456 is clearly offset from the apparent
track traced by the ULIRGs and IR QSOs on this plot.  Therefore,
PDS~456 appears to be a rare, exceptional object, and the duration 
of the physical process governing its present properties must be short
compared with the length of the luminous QSO phase.

\end{abstract}


\section{Introduction \label{sec:intro}}

The remarkable similarity between cosmic star formation
history \citep[e.g.][]{madau96} and quasar evolution
\citep[e.g.][]{shaver96} suggests an intriguing  
possibility of coeval evolution between galaxies and 
massive black holes in the universe.
This is strengthened by the observed correlation
between black hole mass and mass of host galaxy's
spheroid component \citep{magorrian98,ferrarese00,gebhardt00}.
The bulk of star formation and AGN activity in
the early universe may have been driven by mergers of massive,
dusty, and gas-rich galaxies \citep[see][]{blain99}, and
the high frequency of submm continuum detections of high
redshift optically selected QSOs \citep[about 30\%, see][]
{omont01,carilli01} also support the coeval formation and
evolution scenario.  

If the same underlying physical process continues to drive
the formation and evolution of massive galaxies and black holes today,
then evidence for such activities may still be observable.
Citing both starburst and AGN activities
along with large molecular gas contents among the most
luminous infrared galaxies in the local universe, 
\citet{sanders88} suggested an evolutionary connection between
ultraluminous infrared galaxies (ULIRGs) and QSOs.  
Possible formation scenarios for supermassive black holes
inside ULIRGs have been suggested by several theoretical
investigations \citep[][]{weedman83,norman88,taniguchi99}.
Ubiquitous presence of one or more luminous 
active galactic nucleus among ULIRGs has been questioned 
recently \citep[see][]{genzel98,tacconi02}, but
nevertheless a close link between the QSO and the ULIRG 
phenomenon appears to exist and deserves further investigations.

Here, we report a detailed multi-wavelength observational study 
of PDS~456 (IRAS~17254$-$1413), 
the most luminous QSO in the local universe ($z\lesssim0.3$),
in order to shed further light on the ULIRG-QSO connection. 
As part of a systematic survey for young stellar objects at
Pico dos Dias Observatory, \citet{torres97} have obtained
an optical spectrum of the 15th magnitude stellar object PDS~456, selected
based on its bright IRAS detection ($S_{25\mu m}=750$ mJy,
$S_{60\mu m}=930$ mJy).  To their surprise, its optical spectrum
is dominated by broad emission lines characteristic of a QSO
at a redshift of $z=0.184$.  When corrected for
extinction, its absolute B magnitude
of $-26.7$ is 30\% more luminous than 3C~273, the most luminous
quasar in the local universe previously known.
Because it is located near the Galactic Plane ($b=+11^\circ$), 
confusion is a serious
problem at 100 $\mu$m, and only an upper limit of $S_{100\mu m}
\le 1$ Jy is offered by the IRAS data.  Nevertheless, both 
$S_{25\mu m}/S_{60\mu m}=0.81$ and $S_{60\mu m}/S_{100\mu m}\ge 0.93$
suggest PDS~456 is a warm IRAS source, consistent with the
presence of a luminous AGN \citep[see][]{deG85,yun01}.  
Its 60 $\mu$m luminosity alone is 
$\nu L_\nu \sim 10^{12} L_\odot$, high even among the ultraluminous
infrared galaxies in the local universe.  Therefore presence of
significant amount of molecular gas and associated massive
starburst is a strong possibility although not all warm 
ultraluminous IRAS sources are detected in CO emission (see below).  

Given its large bolometric and IR luminosity, PDS~456 represents a
particularly important object, potentially representing 
a critical transition between an ultraluminous infrared galaxy and a
{\it bona fide} QSO with large optical/UV luminosity and broad
emission lines.  We have investigated the nature of the
host galaxy by obtaining a high resolution image
using the Keck telescope in the $K$-band where the QSO-host
contrast may be the most favorable for revealing the underlying
galaxy.  We have also imaged PDS~456 using the VLA in order to
locate the non-thermal activity with subarcsecond accuracy and 
CO (1--0) emission using the OVRO millimeter array
in order to determine whether the host galaxy contains
a large amount of molecular gas, typical of ULIRGs. 
Finally, submillimeter continuum measurements have also been
obtained using the SCUBA camera on James Clerk Maxwell Telescope (JCMT)
in order to constrain the thermal dust spectrum and dust mass.
We examine these new observational results in terms of the
ULIRG-QSO evolution scenario.

\section{Observations \label{sec:obs}}

\subsection{$K$-band Imaging at the Keck telescope \label{sec:Kobs}}

\begin{figure}
\includegraphics[scale=0.75,angle=-90]{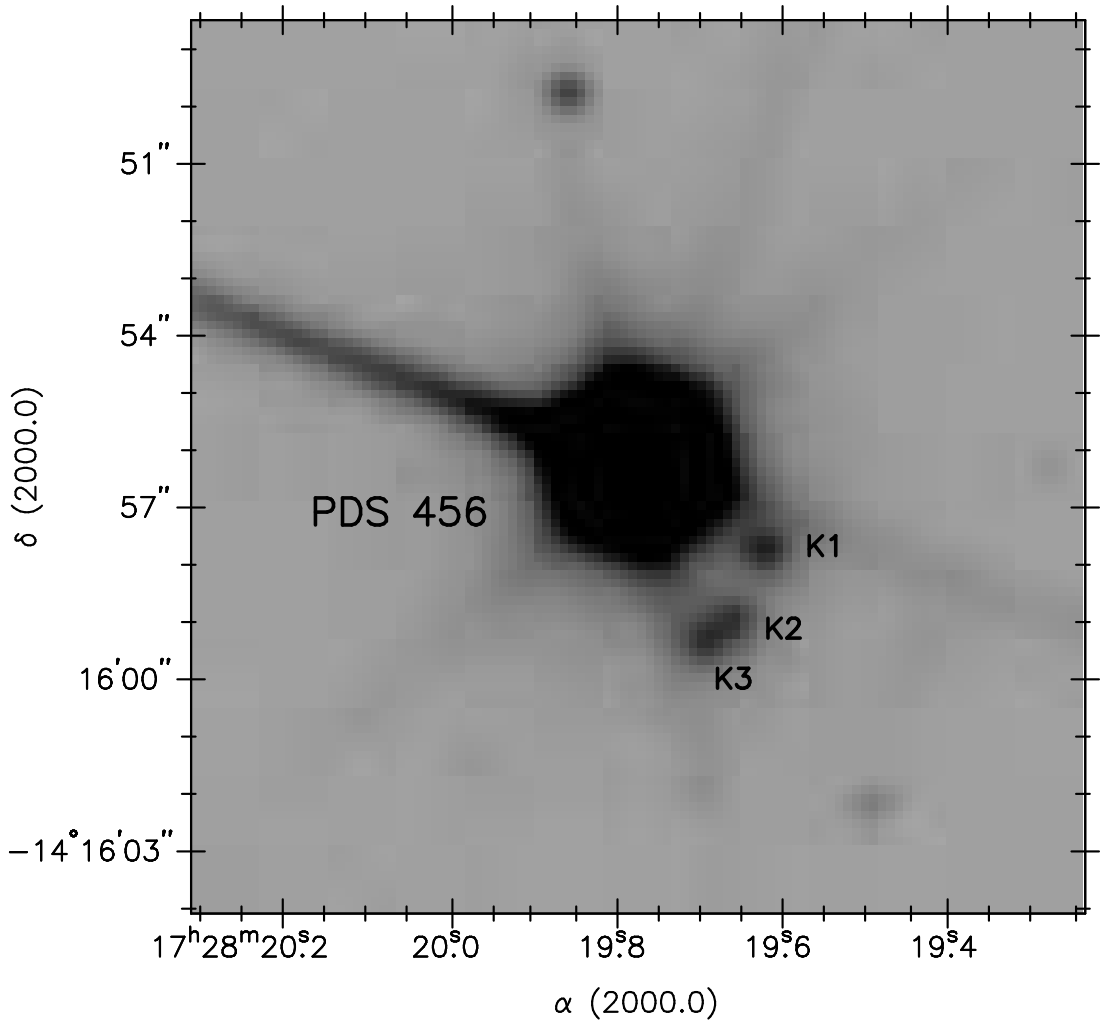} 
\figcaption{Greyscale \kband\ image of PDS~456 obtained at the Keck Telescope.
\label{fig:K}}
\end{figure}

Near infrared observations of PDS~456 were obtained on 
September 1, 2001 with the Near Infrared Camera (NIRC) on the W. M. Keck
telescope \citep{matthews94}.  Observations were carried
out using the $K$ filter centered at $2.21~\mu$m with a FWHM of
$0.43~\mu$m.  The final image was constructed from the 8 dithered
images with 30 second exposure time each, for a total exposure
of 240 seconds.  Seeing was $0.6''$ as measured from several point
sources in the field.  The dithered images were dark-subtracted,
flat-fielded, and then combined using the NIRCtools package written
by D. Thompson and run using the Image Reduction and Analysis
Facility (IRAF).\footnote{
IRAF is written and supported by the IRAF programming
group at the National Optical Astronomy Observatories (NOAO) in Tucson,
Arizona. NOAO is operated by the Association of Universities for Research
in Astronomy (AURA), Inc. under cooperative agreement with the National
Science Foundation.}   
The fractional offsets used to combine
the images were determined from the IRAF task IMALIGN which centroids
on several point sources in the field common to all of the images.
The resulting rotated \kband\ image is shown in Figure~\ref{fig:K}.

\subsection{8.5 GHz and 1.2 GHz Imaging at the VLA \label{sec:vlaobs}}

Radio continuum image of PDS~456 at 8.5 GHz (3.6 cm) wavelength was 
obtained using the Very Large Array (VLA)\footnote
{The National Radio Astronomy Observatory is a facility 
of the National Science Foundation operated under cooperative agreement by 
Associated Universities, Inc.} in Socorro, NM, on January 13, 1998.  
The observations were made in the D-configuration using 14 antennas
in the continuum mode with 100 MHz total bandwidth.
The nearby quasar 1730$-$130 was observed periodically to
track the instrumental gain, and 3C~286 was used for
absolute flux calibration.  The data were calibrated and
imaged following the standard procedure within the NRAO AIPS 
software system.  The naturally weighted image has an
rms noise of $1\sigma=21~\mu$Jy.  PDS~456 is detected with
a peak flux density of $4.5\pm0.9$ mJy ($P_{8.5 GHz}=(3.5\pm0.7)\times 
10^{23}$ W Hz$^{-1}$)\footnote{The quoted 
uncertainty in the flux density reflects the
20\% uncertainty in the transfer of the absolute flux scaling.}, 
unresolved by the synthesized beam of $13\arcsec \times 8\arcsec$ 
(PA=$-29^\circ$) in size. 

Radio continuum image of PDS~456 at 1.2 GHz (25 cm) 
wavelength were obtained using the VLA on July 16, 1999.
The observations were made in the A-configuration using 27 antennas
in the spectral line mode with 6.25 MHz total bandwidth.
The nearby quasar 1730$-$130 was observed periodically to
track the instrumental gain, and 3C~286 was used for
absolute flux calibration.  The naturally weighted image has an
rms noise of $1\sigma=0.26$ mJy.  PDS~456 is detected with
a peak flux density of $24\pm5$ mJy ($P_{1.2 GHz}=(1.9\pm0.4)\times 
10^{24}$ W Hz$^{-1}$), unresolved by the
synthesized beam of $4\arcsec \times 2\arcsec$ (PA=$22^\circ$) in size.
The best astrometric position of PDS~456 is derived from this
image, which is $\alpha(J2000)=17^h~28^m~19^s.78$ and
$\delta(J2000)=-14^\circ~15\arcmin~56\farcs12$ with an
uncertainty of less than $0\farcs4$.

\subsection{CO (1--0) Imaging at OVRO \label{sec:COobs}}

Aperture synthesis CO (1--0) observations of PDS~456
were carried out using the Owens Valley Millimeter Array between
April 2000 and May 2001.  There are six 10.4 m diameter
telescopes in the array, providing a field of view of about $75''$
(FWHM) at 100 GHz.  The telescopes are equipped with SIS receivers
cooled to 4 K, and typical single sideband system temperatures 
($T_{SSB}$) were between 250 and 350 K.  
Baselines of 15-115 m E-W and 15-115 m N-S were used.  
The total on-source integration time was about 28 hours.  

The redshifted CO (1--0) line at $z=0.184$ occurs at 97.3574 GHz.
A digital correlator configured with $120 \times 4$ MHz
channels (12 \kms) covered a total velocity range of 1440 \kms.
By tuning the receivers to this frequency, the spectrometer
covers the redshift range of $z= 0.1811 - 0.1869$.
The nearby quasar 1730$-$130 was observed at 20 minute intervals to
track the phase and short term instrument gain.  Uranus ($T_B=120$ K),
Neptune ($T_B=115$ K), 3C~273, and 3C~454.3 were observed for the
absolute flux calibration.  The data were calibrated using the
standard Owens Valley array program MMA \citep{Sco92} and were mapped
and analyzed using the imaging program DIFMAP \citep{Shepherd94} and
the NRAO AIPS software system.  The uncertainty in absolute flux
calibration is about 20\%, mainly due to the uncertainty in
transferring the calibration between the source and the flux calibrators.
The positional accuracy of the
resulting maps is better than $\sim 0\farcs5$.

\subsection{850 $\mu$m Photometry \label{sec:SCUBAobs}}

The 850 $\mu$m photometry observations were conducted on 
October 26 \& 29, 2002, using the submillimeter camera SCUBA on 
the James Clerk Maxwell Telescope (JCMT) in a photometry mode. 
The 225 GHz zenith opacity was 0.1 and 0.06, respectively, on 
those days, and high temporal resolution measurements from the 
JCMT line-of-sight water-vapor radiometer were used to correct 
for the opacity variations. 
The flux calibration was done using Uranus.
Due to the early times in the night, the atmosphere was somewhat 
unstable.  This was compounded by the low elevation ($EL\sim 30^\circ$)
of the source at the time of the observations.
The sensitivity achieved is about 2.5 mJy ($1\sigma$) after a
total of 2 hours of integration.  The data does not reveal any
positive emission above the noise level, but the 3$\sigma$
upper limit of 7.5 mJy at 850 $\mu$m provides a useful constraint 
on the dust continuum spectrum and the amount of cold dust
associated with the host galaxy of PDS~456.

\section{Results \label{sec:results}}

\subsection{$K$-band Image \label{sec:K-results}}

The final co-added \kband\ image of PDS~456 is shown in 
Figure~\ref{fig:K}.  The point-like QSO and the 
point spread function (PSF) of the telescope dominates the 
image, and three compact sources lying about  $3''$ to the southwest 
of the QSO are seen (named K1, K2, and K3).  
We performed aperture photometry on the QSO as well as the three
compact components, and the results are summarized in Table~\ref{tab:K}.
Photometric zero-points are determined using separate
observations of several standard stars observed at varying airmass
throughout the night.  Aperture
corrections were made for the $5$, $7$, and $20$ pixel radius aperture
for K1, K2+K3, and the QSO, respectively.  We 
derive \kband\ magnitude of $9.89\pm 0.02$ for the QSO.
After subtracting the wings of the QSO, we derive a \kband\ magnitude
of $16.3\pm 0.2$ for K1.  K2 and K3 are approximately equal in
brightness, and the combined \kband\ magnitude is $15.6\pm 0.1$. 
The near-IR photometry for the QSO reported by the Two Micron All Sky 
Survey \citep[2MASS: ][]{skrutskie97} 
is $m_J=12.03\pm0.02$, $m_H=11.08\pm0.02$, and 
$m_K=9.86\pm0.02$ (see Table~\ref{tab:sed}), in an excellent
agreement with our measurement.  The absolute astrometry
for the \kband\ image is good to only about 1\arcsec, and 
the astrometric information is improved by replacing the
the \kband\ QSO centroid position with the radio centroid position 
(\S~\ref{sec:vlaobs}).

\begin{figure}
\plotone{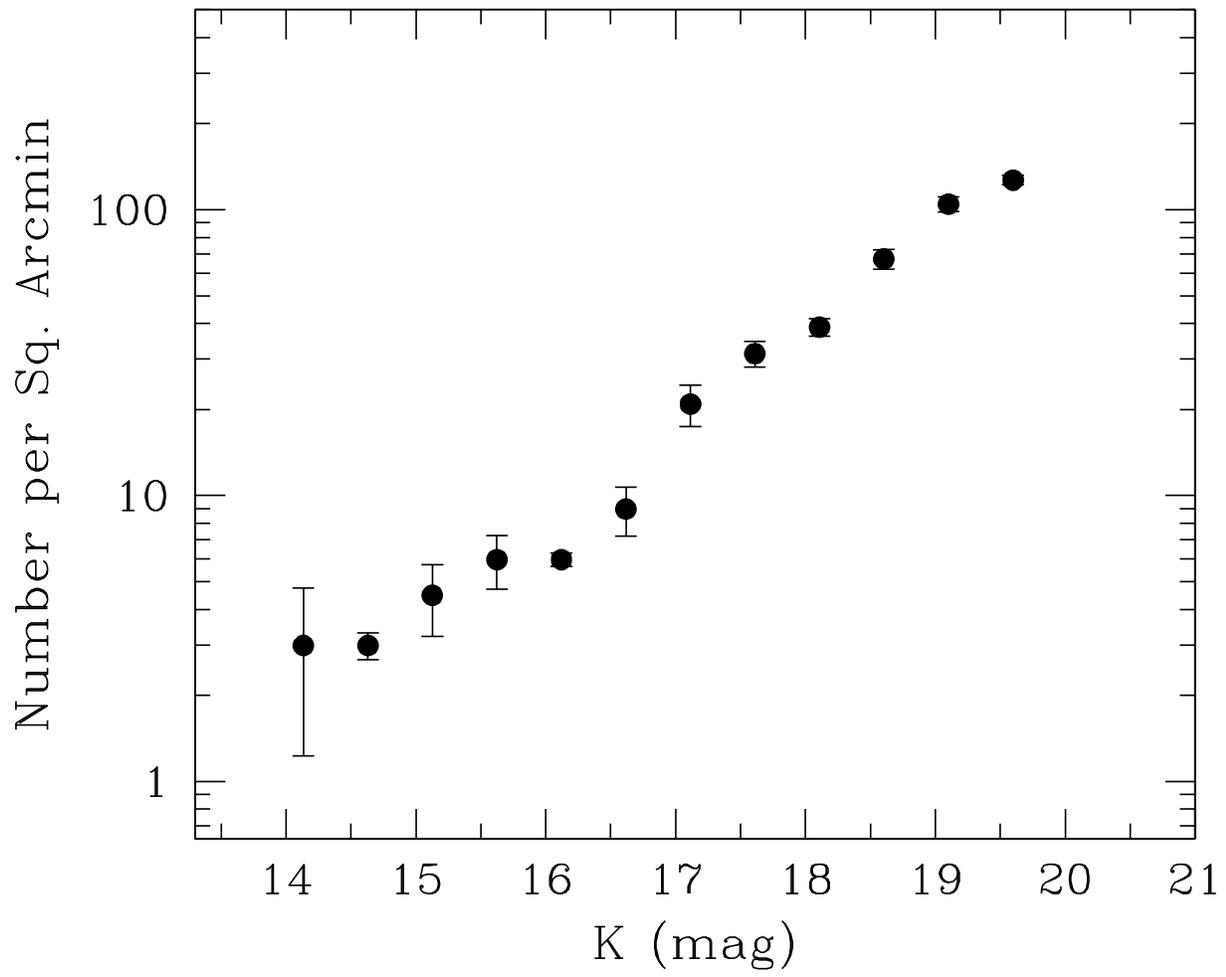}
\figcaption{Cumulative number counts of $K$-band sources
in the PDS~456 field.  Stars dominate the brightest sources.
\label{fig:Kcounts}}
\end{figure}

The whole co-added \kband\ image is about 50\arcsec\ by 50\arcsec\ in
size and includes about 90 sources brighter than $K=20$.  
The integrated source count
is shown in Figure~\ref{fig:Kcounts}.  The number density of
sources brighter than $K=16.5$ is about 10 per square arcminute, and
this corresponds to about $3\times 10^4$ objects per square degree, 
which is
nearly two orders of magnitudes higher than the \kband\ galaxy
count in the field \citep[e.g., see Fig.~2 by ][]{djorg95}.
A natural explanation for the elevated source count is that
PDS~456 is located in the direction of Galactic Center,
near the Galactic Plane, at $l=10^\circ.4$ and $b=+11.^\circ2$.
Therefore foreground stars likely account for most of the sources 
found in the \kband\ image.

The three compact features K1, K2, and K3 found adjacent to the
QSO are very likely physically associated with PDS~456, and the host 
galaxy may have undergone a tidal interaction or a merger recently.
For $H_\circ=75$ km s$^{-1}$ Mpc$^{-1}$, $\Omega_m = 0.3$, and 
$\Omega_\lambda = 0.7$,
angular distance for PDS~456 at $z=0.185$ is 595 Mpc, and the
angular separation of 3\arcsec\ corresponds to a projected distance
of 9 kpc, which is the natural size scale for the underlying host 
galaxy or companions.  
The source count shown in Figure~\ref{fig:Kcounts}
suggests that the probability of a random $K=16.5$ source
falling within a 3\arcsec\ radius of any given position is less
than 1\%.  The likelihood of a chance projection for 
{\it three} random sources of this magnitude within 
3\arcsec\ of PDS~456 is less than $10^{-6}$.  Therefore, a random
occurrence of three foreground or background sources near the
position of PDS~456 is extremely unlikely.
If they are at the same distance
as PDS~456, their \kband\ absolute magnitude ($M_K$) ranges between
$-20.7$ and $-21.1$, which is about 4 magnitudes fainter than that of
an $L^*$ galaxy.  Attempts to subtract the QSO image
using the PSF derived from calibration stars were not
successful in revealing any underlying structures 
within $\sim4$ kpc of the QSO ($M_K=-27.5$).   
An L$^*$ stellar host may be present underneath the QSO image,
but it cannot be determined by our present data.

A faint extension to the southwest of the QSO is also
visible even in the Digitized Sky Survey and 2MASS
plate images.  This feature is not clearly resolved by these 
low resolution images, however, and little
else can be learned from them.  In an unpublished $J$-band
image obtained with the Pico dos Dias Observatory 1.6 m telescope 
in 1999 by F. Jablonski under a good seeing condition, 
sources K1 and K2+K3 appear clearly resolved, and the K1 source, which 
appears extended toward the QSO, has a $J$-band magnitude of
$17.8\pm0.2$ while the combined brightness of K2+K3 is $17.5\pm0.1$ 
(F. Jabloski \& C. Torres 2003, private communication).  
The $J-K$ colors of K1 and K2+K3, $1.5\pm0.3$ and $1.9\pm0.2$
respectively, are somewhat redder than the ensemble average color of 
2MASS galaxies, $<J-K>= 1.2$, but they are consistent with that 
of a stellar system located at the redshift of the QSO when the 
foreground extinction (see \S~\ref{geometry}) and the 
$k$-correction are taken into account.  In an imaging survey of
20 nearby luminous QSOs at $z<0.3$ using the HST, \citet{bahcall97}
reported finding eight companion galaxies with $V$-band magnitude
within 4 magnitudes of $L^*$ at projected distances less than
10 kpc.  Other HST imaging studies have also reported a high
frequency of tidal tails and close companions associated with 
the nearby luminous QSOs \citep{mcleod01,dunlop03}.  Therefore
finding compact stellar systems surrounding PDS~456 itself
is not unusual, but the large number and the asymmetric distribution 
are curious.  The asymmetry and the clumpy appearances are
similar to the features seen around 3C~273 \citep{martel03} and
PKS~2349$-$014 \citep{bahcall97,mclure99}, and therefore 
these stellar clumps are also likely the remnants of a recent 
accretion or a merger event.  A definitive characterization of
the nature of these clumps will require a more sensitive 
higher angular resolution imaging using the HST.

\subsection{CO (1--0) and 100 GHz Continuum \label{sec:CO-results}}

\begin{figure}
\plotone{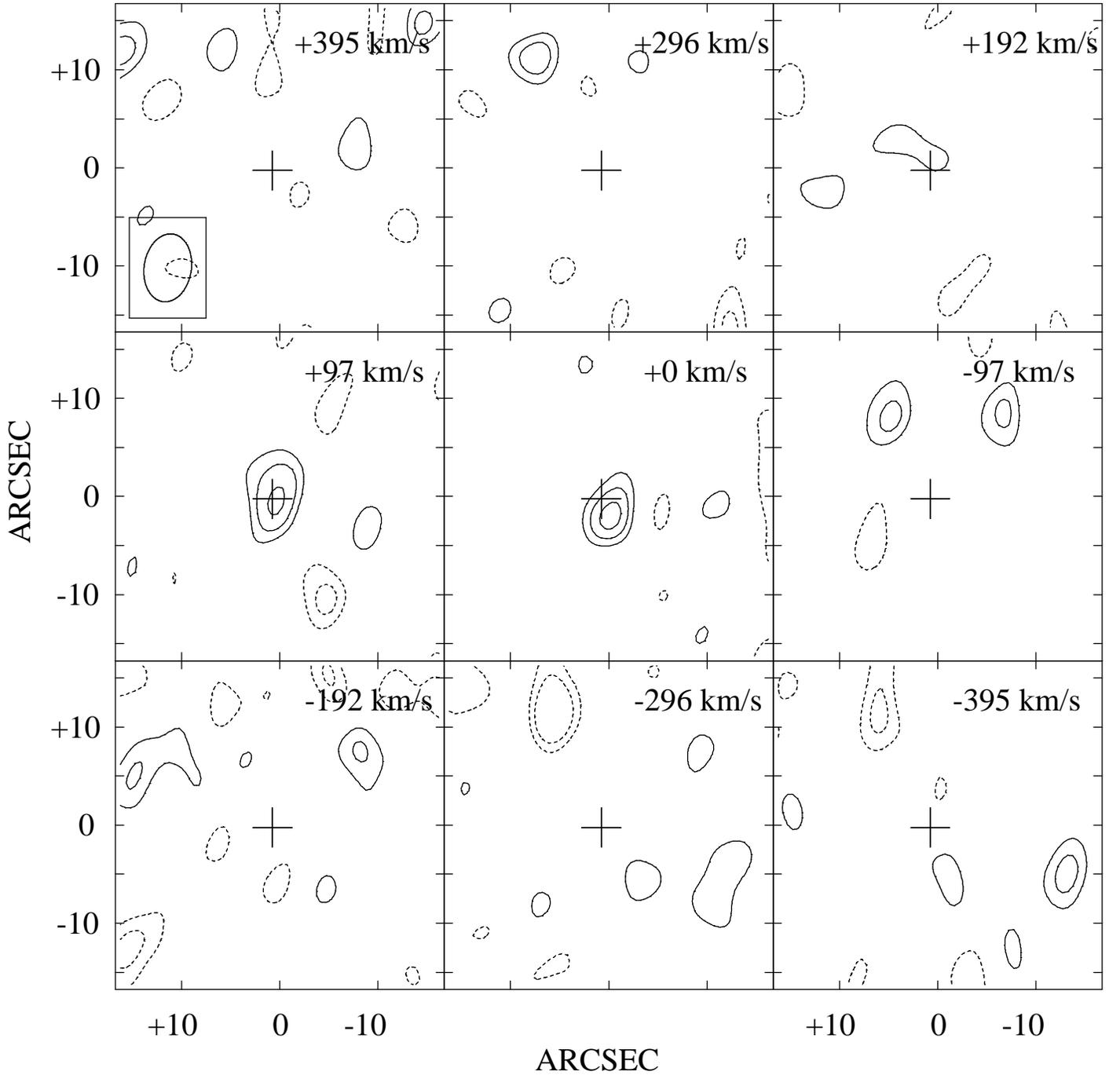}
\figcaption{CO (1--0) channel map of PDS~456.  The contour levels correspond
to $-3$, $-2$, +2, +3, +4, and +5 times 1.5 mJy beam$^{-1}$ ($1\sigma$).
The velocity offsets listed are with respect to the channel with the
strongest line feature at 97.301 GHz.  The cross marks the optical
QSO position.
\label{fig:chmap}}
\end{figure}

\begin{figure}
\plotone{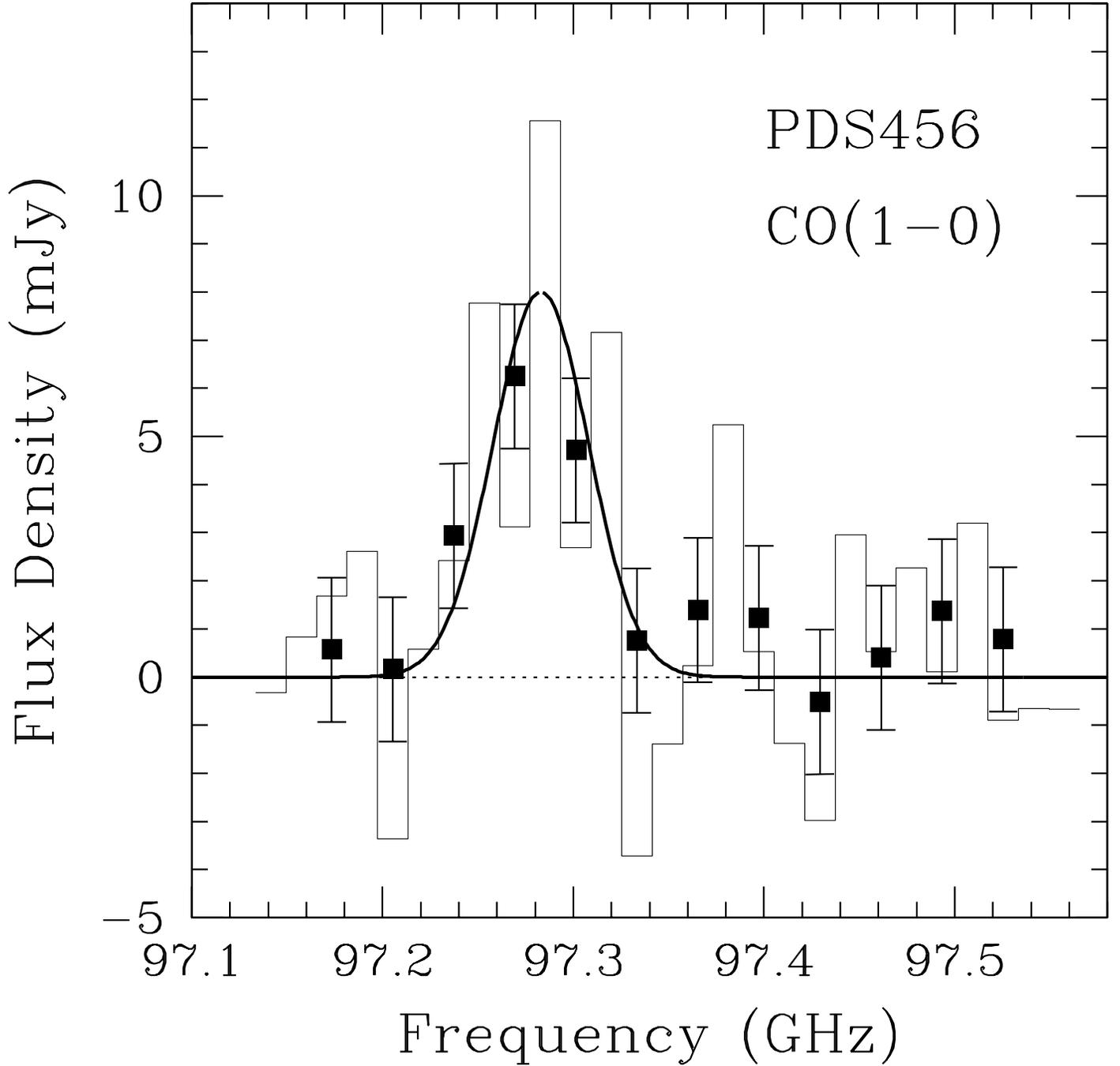}
\figcaption{CO (1--0) spectrum of PDS~456.  The histogram is the spectrum
obtained at the QSO position after smoothing to the 16 MHz (49 \kms)
resolution.  The CO spectrum further smoothed to a 64 MHz (197 \kms) 
resolution is shown using filled squares with an increment of
32 MHz.  The thick solid curve is a
model Gaussian line profile with $\sigma = 25$ MHz
(77 \kms; FWHM = 181 \kms) centered at 97.283 GHz ($z=0.1849$).
\label{fig:spectrum}}
\end{figure}

The CO (1--0) channel maps of PDS~456 are shown in Figure~\ref{fig:chmap}. 
The naturally weighted image has a synthesized beam of about $7.0\arcsec
\times 4.8\arcsec$ (PA=$-9^\circ$). 
Emission features brighter than 6 mJy beam$^{-1}$ (4$\sigma$)  
are seen within 3\arcsec\ of the QSO position marked with 
a cross at channels corresponding to 97.269 GHz and 97.301 GHz.
The CO (1--0) spectrum obtained at the QSO position after 
smoothing to the 16 MHz (49 \kms) and 64 MHz (197 \kms) 
resolution is shown in Figure~\ref{fig:spectrum}.  The thick 
solid curve is a model Gaussian line profile with $\sigma = 25$ MHz
(77 \kms; FWHM = 181 \kms) centered at 97.283 GHz ($z=0.1849$).
The velocity integrated line flux is $1.45\pm0.29$ Jy \kms. 
At a luminosity distance of 834 Mpc, 
this translates to a H$_2$ mass of $9\times 10^9 M_\odot$ using the
standard conversion factor derived from the Galactic molecular 
clouds \citep[see][]{young91}.  This derived H$_2$ mass is
on the high end of the mass range found for PG QSOs
\citep[1-10$\times 10^{9} M_\odot$, ][]{evans01,scoville03} and
is only slightly less than those of ULIRGs \citep[2-5$\times 10^{10} 
M_\odot$ using the same conversion, e.g. ][]{solomon97}.

The spatial resolution of the CO data is about 21 kpc $\times$ 15 kpc,
and the CO emitting structure is not expected to be substantially 
resolved.  However, the centroids of the brightest CO peaks in 
the two channel maps shown in Figure~\ref{fig:chmap} 
are offset from each other by about 2\arcsec\ (6 kpc).  
Given the low S/N of the data, this displacement
of the peak positions may not be very significant.  On the other hand,
the displacement of the blueshifted CO peak is in the direction of
the two \kband\ sources K2 \& K3 (see Fig.~\ref{fig:K}).
These two \kband\ sources are likely associated with the host of 
the QSO (see \S~\ref{sec:K-results}), and this apparent alignment
with the CO displacement, 
perhaps fortuitous, adds to the possibility that the molecular gas 
associated with PDS~456 may be extended over 5-10 kpc scales.
 
The data from the image sideband of the OVRO receivers are also
cross-correlated to produce continuum visibility data centered
at 100.3574 GHz.  The 100 GHz continuum image obtained from 
these visibility data has an rms noise of 0.2 mJy beam$^{-1}$,
and no statistically significant source is found with the 
field of view of the interferometer.  From this we can place 
an upper limit of 0.6 mJy ($3\sigma$) for the QSO continuum 
flux density at 100 GHz.  The continuum spectral energy 
distribution of PDS~456 is discussed in greater detail below.

\section{Discussions \label{sec:discussions}}

\subsection{Spectral Energy Distribution of PDS~456 \label{sec:sed}}

The spectral energy distribution (SED) holds information 
on the temperature and the physical processes involved, even when the 
angular resolution of the observations are insufficient to yield
any structural information.  In examining the ULIRG-QSO connection,
a comparison of the SEDs is particularly useful since
the underlying physical processes and host galaxy properties
can be directly compared.

\begin{figure}
\plotone{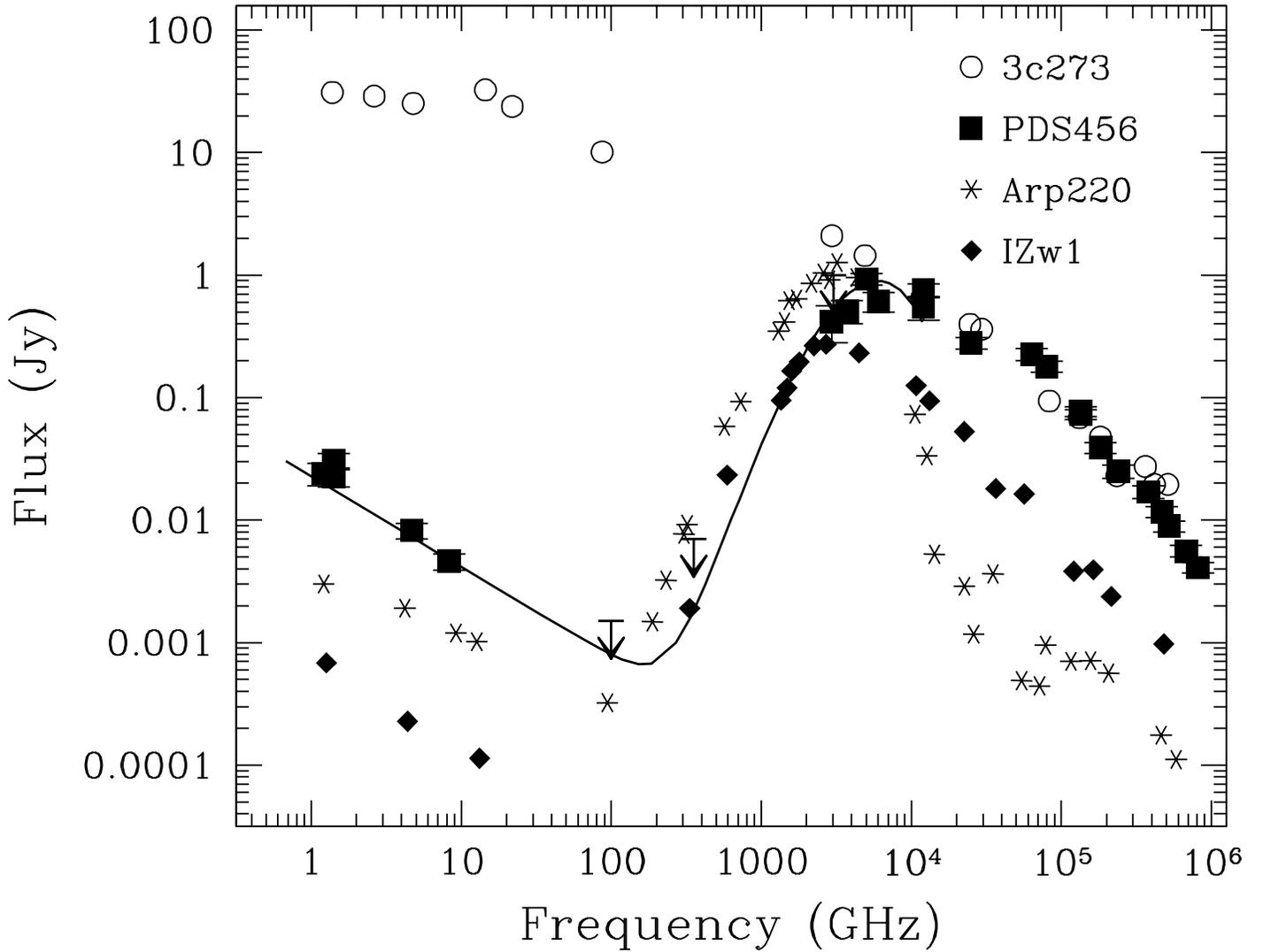}
\figcaption{Spectral Energy Distribution (SED) of PDS~456.
The SEDs of 3C~273 ($z=0.158$), I~Zw~1 ($z=0.061$), and 
Arp~220 ($z=0.018$) are also
shown for comparison after redshifting to that of PDS~456
($z=0.185$).  The solid line is a model SED for a source with
dust temperature of $T_d=120$ K, emissivity $\beta=1.5$, and
a FIR emission region with an effective diameter of 135 pc
\citep[see ][]{yun02}.
The radio continuum part of the model SED is a power-law
with a slope of $\alpha=+0.75$, and the flux density is boosted
by a factor of eight over the value predicted by the radio-FIR 
correlation for star forming galaxies in order to
match the observed data points.
\label{fig:sed}}
\end{figure}

The SED of PDS~456, shown with large squares in Figure~\ref{fig:sed},
nicely demonstrates its QSO-like and ULIRG-like nature.
The SEDs of the previously most luminous QSO 3C~273 ($z=0.158$),
an IR luminous radio-quiet QSO I~Zw~1 ($z=0.061$), and the
prototypical ULIRG Arp~220 ($z=0.018$) are shown after 
redshifting their continuum spectra to $z=0.185$ for a direct
comparison.  The SED data for PDS~456 are also tabulated in
Table~\ref{tab:sed}.  The SED data for 3C~273, I~Zw~1, and Arp~220
come from the NASA Extragalactic Database (NED) and literature.
The SEDs of PDS~456 and 3C~273 are remarkably similar and are 
essentially identical between the far-IR (FIR) and optical wavelengths.
When corrected for extinction of $A_V\sim 1.5$, the bolometric 
luminosity of PDS~456 becomes about 30\% larger than that of 3C~273 
\citep{torres97,simpson99}. 
Identified as one of the ``warm ULIRGs'' that are possibly
in transition from galaxy to quasar by \citet{sanders88b},
3C~273 has both thermal and non-thermal emission mechanisms
contributing to its SED \citep[see][]{courvoisier98}.
The SEDs at radio wavelengths differ by more than 3 orders of
magnitude, reflecting the presence of a beamed radio jet in 3C~273.
This part of the SED contributes little to the total luminosity,
however.  

The SED of PDS~456 covering the far-IR through radio wavelengths
is modeled using the dusty starburst SED model by
\citet{yun02} and is shown using a solid curve in Figure~\ref{fig:sed}.  
As in QSOs 3C~273 and I~Zw~1, non-thermal emission from the AGN dominates 
the short wavelength ($\lambda < 50 \micron$) part of the SED in PDS~456, 
but dust reprocessed AGN emission and light from young stars should 
account for the dust peak from millimeter to FIR wavelengths 
\citep[e.g.][]{rowan95}.
There are no continuum detections of PDS~456, only upper limits,
between 100 GHz (3000 $\mu$m) and 2000 GHz (150 $\mu$m),
and therefore there is only limited information to constrain the cold
gas content and cold dust properties.  The turnover
in the FIR dust peak occurs at a much higher frequency, near
6000 GHz (50 $\mu$m), implying the dust temperature 
is on average much higher than that in Arp~220
\citep[$T_d\sim 45$ K, ][]{scoville91,yun02}.  The model 
SED shown in Figure~\ref{fig:sed} is chosen to match the 
100 $\mu$m measurement by \citet[][]{reeves00} and has dust temperature 
of 120~K in its rest frame.  This means a substantial amount of
dust in the host galaxy of PDS~456 is exposed to a significantly
higher mean radiation field and to the high energy
photons originating from the AGN.  The total dust mass derived
from the existing SED data, assuming $T_d=120$ K and emissivity
$\beta=1.5$, 
is $0.4 \times 10^7 M_\odot$ \citep[using Eq.~3  of][]{yun98}. 
Both the dust mass and temperature are well within the range of
dust properties derived for a large sample of Palomar-Green (PG)
QSOs by \citet[][ $T_d=20-120~K,~M_d=10^{7\pm1} M_\odot$]{Haas00}. 
The effective diameter for the emitting area is about 135 pc 
\citep[using Eq.~4 of][]{yun02}, which is similar in size to the
nuclear starburst region in Arp~220 \citep[][]{sco97,sakamoto99}.
The strong mid-IR 
emission from PDS~456 and 3C~273 are indicative of abundant
warm and hot dust (a few hundred to 1000 K), and this is 
a clear indication of an energetic AGN in these objects 
\citep[see Figure~10 of][]{yun01}.  
The measured radio continuum flux density of PDS~456 is 8 times larger 
than what is expected from a starburst system with the same 
FIR luminosity, and this is a 
clear indication that a radio AGN is also present in PDS~456.

The total dust mass of PDS~456 could be potentially much larger if 
substantial amount of cold ($T_d\sim$ 10-40 K) dust is present.
The best constraint on the cold dust content come
from our new 850$\mu$m measurement made using the SCUBA camera
on JCMT.  As shown in Figure~\ref{fig:sed}, the dust continuum
emission from Arp~220 should have been detected by 
our observation with S/N $\gtrsim 5$.  The $3\sigma$ upper 
limit at 850 $\mu$m shown here suggests that the total cold dust
mass of PDS~456 is less than one half of the amount in Arp~220
($T_d\sim 45$ K).  The model SED for PDS~456 shown in 
Figure~\ref{fig:sed} is consistent with that of the composite
starburst+QSO system I~Zw~1 with estimated cold dust mass of $(1-6)\times
10^7 M_\odot$ \citep{hughes93,haas98,andreani99}.  More 
sensitive submillimeter continuum measurements are needed
to constrain the total dust mass in PDS~456.

\begin{figure}
\plotone{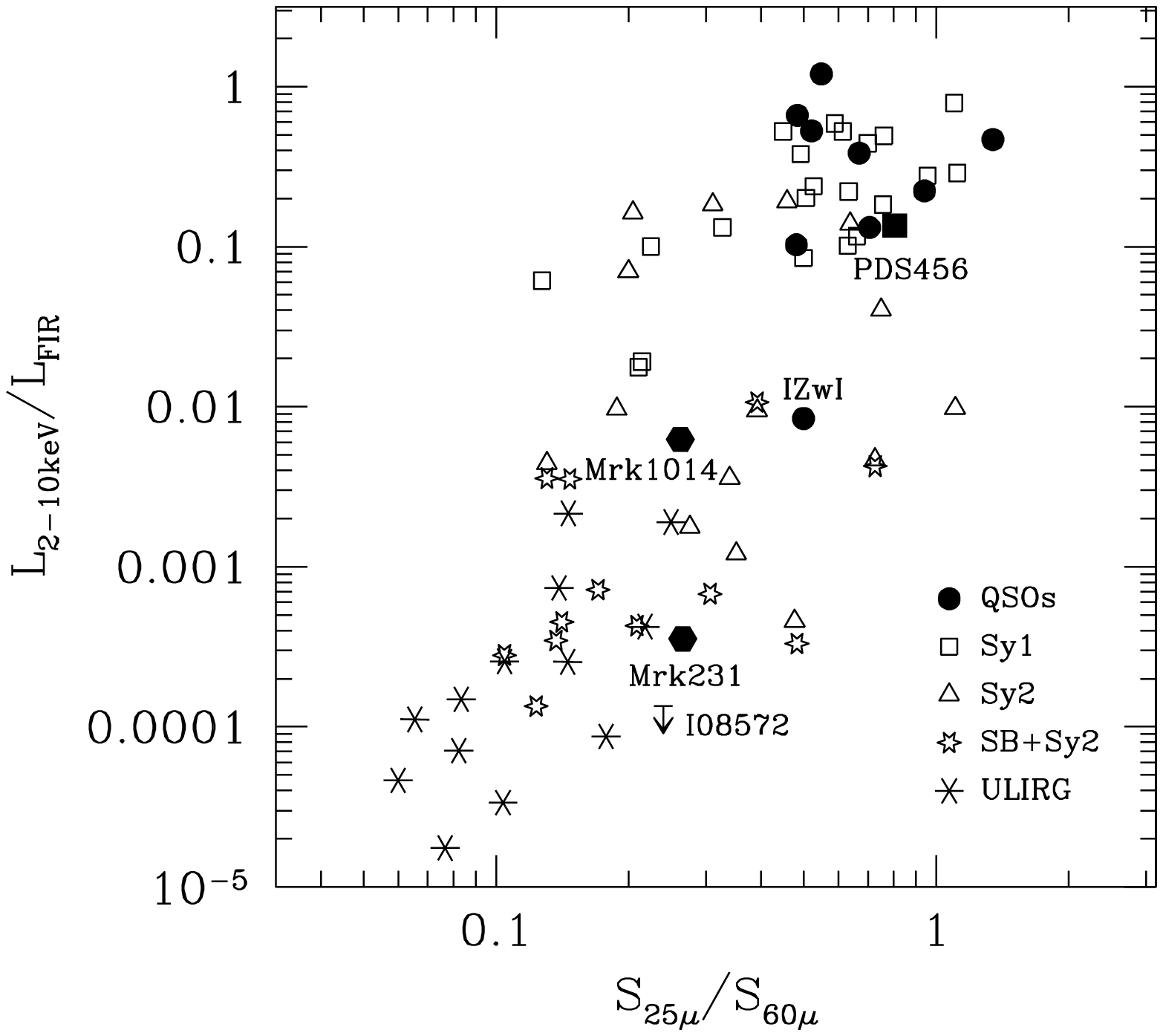}
\figcaption{$L_{2-10kev}/L_{FIR}$ vs. $S_{25\mu}/S_{60\mu}$.
The X-ray data plotted come from \citet{mas-hesse95}, 
\citet{vignali00}, \citet{risaliti00},
\citet{levenson01}, and \citet{ptak03}.  Three luminous
IR QSOs Mrk~231, Mrk~1014, and I~Zw~1 are identified along
with PDS~456 for comparison.  The warm IRAS source
IRAS~08572+3915 is not detected in hard X-ray 
\citep[$L_{2-10kev}<4.4\times 10^{41}$ erg s$^{-1}$, ][]{risaliti00},
and an upper limit is shown.
\label{fig:xray}}
\end{figure}

A useful insight into the AGN contribution to the observed FIR
and bolometric luminosity can be obtained by examining
its hard X-ray luminosity.  The best-fit model spectrum for the BeppoSAX
measurements yields an intrinsic 2-10 keV luminosity of about
$5.3\times 10^{44}$ erg s$^{-1}$ \citep[][after correcting for 
$H_\circ$]{vignali00}, which is about 10\% of its
total FIR luminosity, $L_{FIR}=1.3\times 10^{12} L_\odot$.
If the AGN activity should account for a substantial fraction of 
its FIR and bolometric luminosity, one would expect the X-ray luminosity of
the AGN to be comparable or larger than the FIR luminosity.
While the intrinsic 2-10 keV luminosity does not account for the
entire FIR luminosity, this fraction is quite substantial,
similar to other IR bright, optically selected QSOs and Seyfert 1 AGNs and 
and 2-3 orders of magnitudes larger than most
ULIRGs and composite starburst+AGN systems \citep[see
Figure~\ref{fig:xray}; also ][]{risaliti00,levenson01}.  PDS~456
stands out even among the FIR bright QSOs as its
$L_{2-10 keV}/L_{FIR}$ ratio is an order of magnitude
larger than that of IR luminous QSOs Mrk~1014 and I~Zw~1
and two orders of magnitudes larger than that of a prototypical
ULIRG/QSO system Mrk~231.  Therefore, the hard X-ray properties
of PDS~456 is much closer to optically selected luminous QSOs,
and it can be strongly argued that the bolometric 
luminosity of PDS~456 is dominated by the AGN activity.

\subsection{Geometry of the Circum-AGN Disk  \label{geometry}}

Most of the warm dust responsible for the mid- and far-IR 
emission modeled in Figure~\ref{fig:sed} has to be located within 
a rotationally supported circum-AGN disk surrounding the 
central AGN in order to achieve the observed
high mean temperature.  A {\it lower} limit to the 
dynamical mass of the gas/dust disk surrounding the optical
QSO in PDS~456 can be estimated using the FIR source size 
derived in \S~\ref{sec:sed} and the gas rotation speed 
derived from the CO line width in Figure~\ref{fig:spectrum}.
Assuming the gas is in a circular rotation with rotation speed
of $V_c=90~(sin~i)^{-1}$ \kms\ at a radius of 70 pc,
the resulting dynamical mass is
$$M_{dyn}=1.3\times 10^8 ~[{{R}\over{70~pc}}]
[{{V_c}\over{90~km/s}}]^2(sin~i)^{-2} M_\odot.$$
This is far smaller than the gas mass inferred from the CO luminosity,
$9\times 10^9 M_\odot$ (see \S~\ref{sec:CO-results}), unless the
gas disk is viewed nearly face-on with an inclination angle smaller 
than $i\lesssim 8^\circ$.  The CO-to-H$_2$ conversion
factor is thought to be 2-3 times smaller than the standard
Galactic value for the warm, dense clouds in circum-nuclear regions 
\citep{sco97,downes98} while both $V_c$ and $R$ are only known to 
a factor of two or so.  However, these factors alone cannot fully 
account for the difference between the inferred molecular gas mass and the 
dynamical mass.  

The simple fact that the optical QSO is readily visible requires 
the line-of-sight to the QSO be mostly free of any obscuring 
material, and this also favors a face-on geometry for the circum-AGN disk.
A similar inference on the face-on nature and the resolution of
the apparent discrepancy between the derived gas mass and the 
dynamical mass for gas disks within
ULIRGs hosting an optical QSO (e.g. Mrk~231) has been made 
previously \citep{bryant96,downes98}.  Since PDS~456 is seen through
the Galactic plane, some foreground extinction is
naturally expected, and the spectral slope between 
the near-IR and optical wavelengths is indeed steeper than expected.  
The analysis of the optical and near-IR colors, continuum 
spectral shape, and the Balmer line ratios all consistently 
suggests reddening of the QSO light by $A_V\sim 1.5$ 
\citep[see ][]{torres97,simpson99}.  Torres et al. further
estimate that the reddening due to the Galactic foreground 
derived from the extinction map of \citet{burstein82} and 
from the diffuse interstellar bands and Na D1 line in their 
spectra also suggest a foreground reddening of $A_V\sim 1.5$, 
accounting for {\it all} of the reddening associated with the QSO light. 
The only result that contradicts this conclusion
is the analysis of the X-ray spectrum that requires a highly ionized
absorber with $N_H(warm) \sim 5\times 10^{24}$ cm$^{-2}$ and an
additional cold absorber with $N_H(cold) \sim 3\times 10^{22}$
cm$^{-2}$ \citep{vignali00}.  However, this inference is highly
model dependent.  
 
The CO emission from PDS~456 may instead arise primarily from the
cold gas and dust yet undetected in continuum, rather than from the
circum-AGN disk traced in IR in Figure~\ref{fig:sed}.  
The channel maps shown in Figure~\ref{fig:chmap} suggests that the
CO emitting region may be extended over a region 10 kpc in extent
(see \S~\ref{sec:CO-results}).  Even with the physical extent
of 5-10 kpc for the CO emitting region, 
the dynamical mass $M_{dyn}$ is still smaller than the H$_2$
mass inferred from the CO luminosity unless the inclination
of the disk is more face-on ($i\lesssim 30^\circ$).  

\begin{figure}
\plotone{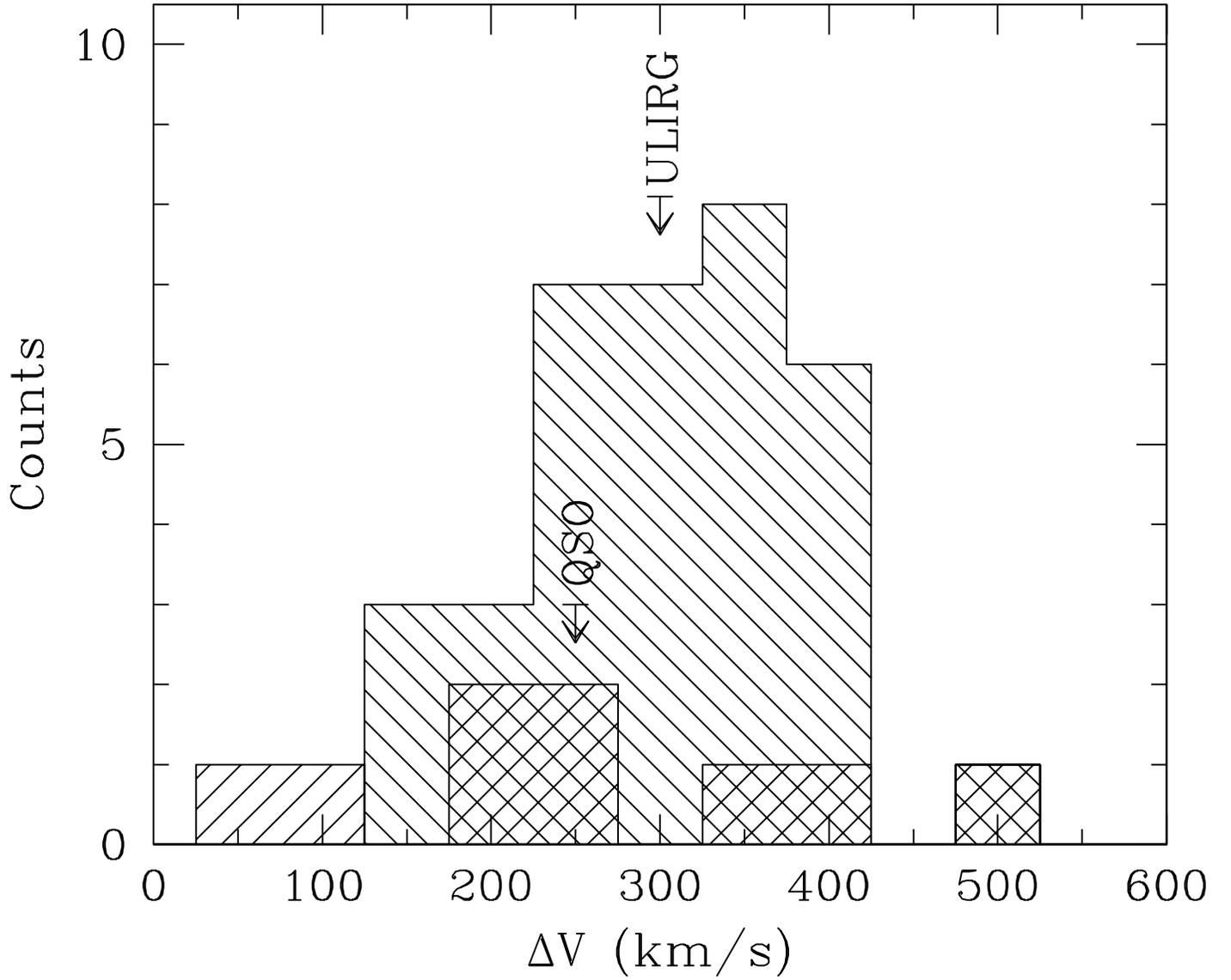}
\figcaption{A histogram of CO line widths for the optically
selected QSOs and ULIRGs.  The sources plotted are the same as
in Figure~\ref{fig:SFE}.  The median line widths for the QSOs
and the ULIRGs are also marked.
\label{fig:linewidth}}
\end{figure}

While the face-on geometry for the dust and gas disks in PDS~456 
and Mrk~231 seems quite secure, a surprising result is that
a general inference of a face-on geometry for nuclear gas disks 
in other QSO host systems is {\it not} supported by
the observed distribution CO line widths.  As demonstrated by the
discussion above, the face-on geometry requirement arises mainly 
from the small observed line widths in the 
dynamical mass calculation.  However, when the histograms
of observed CO line widths are compared between QSO
hosts and ULIRGs as shown in Figure~\ref{fig:linewidth},
QSO hosts show comparable CO line widths as ULIRGs.
The median CO line width for ULIRGs is larger 
($\Delta V\sim 300$ \kms) than that of QSO hosts 
($\Delta V\sim 250$ \kms), consistent with the general expectation,
but the difference is much smaller than expected from the viewing
geometry consideration alone.  
Furthermore, the CO line widths observed in three out 
of nine QSO hosts are larger than 350 \kms\ (FWHM), too large to 
be consistent with a face-on geometry.  One solution to this
general problem is that the parsec scale circum-AGN disk that dictates the
viewing angle of the central source is not aligned with the 
extended gas/dust disk or ring traced in CO. Since only a few of
these objects have been imaged in CO or in dust continuum 
with sub-kpc resolution, the possibility of optical AGNs
viewed through gaps in a large scale gas/dust ring or disk at an 
arbitrary inclination angle or through a highly warped disk cannot be
ruled out.  Therefore, Figure~\ref{fig:linewidth} cautions 
strongly against the common assumption of 
a face-on geometry for gas and dust distribution around optically 
visible QSOs.,

\subsection{Dusty QSO hosts and ULIRGs \label{comparison}}

In the ULIRG-to-QSO evolutionary scenario proposed by
\citet[][]{sanders88}, an ultraluminous infrared galaxy is
thought to go
through a warm IR source phase as observed in PDS~456 today
with large IR luminosity, warm IR color, a large molecular
gas content, and a dust enshrouded luminous AGN.  In accordance
with this scenario, a large number of optically selected PG
quasars that are also detected by IRAS have been studied
in detail and are found to contain large amounts of molecular
gas \citep{alloin92,evans01,scoville03}.  Similarly, the majority of 
warm ULIRGs are also shown to be composite objects hosting one or
more luminous AGNs and an active starburst fueled by the large
molecular gas content in their nuclear regions \citep{sanders88b}.

IR luminosity and total gas/dust mass are the two key variables 
characterizing the ULIRG-QSO  evolution.   
Specifically, the transition from a ULIRG to a QSO
is driven by rapid conversion of gas into stars and/or
the subsequent growth of a massive black hole, followed by
the dispersion of gas and dust surrounding these activities.
Therefore such an evolution scenario should follow a distinct track
in the plot between IR luminosity and the total gas mass.
In particular, a dusty QSO emerging from 
a ULIRG phase should show the characteristic large FIR luminosity
and rapidly diminishing gas content, distinguishable from 
ULIRGs.   If the IR-bright
QSO PDS~456 is such a transition object, then it should
appear in the bridging region between the areas occupied by
ULIRGs and by QSOs. 

To scrutinize the ULIRG-QSO evolutionary scenario further, 
optically selected IR QSOs, including PDS~456, are compared 
with ULIRGs and luminous IR galaxies (LIRGs) in a plot
of FIR luminosity ($L_{FIR}$) versus CO luminosity ($L'_{CO}$), 
which can be translated into a total molecular gas mass ($M_{H_2}$) 
using the standard conversion relation \citep[see ][]{young91}.  
This plot shown in Figure~\ref{fig:SFE} is  
commonly referred to as the ``star formation efficiency (SFE)'' plot 
because historically it is used to demonstrate that IR-bright
starburst systems are not only forming stars at high rates 
but also with a greater efficiency, producing 1-2 orders of 
magnitudes more luminosity (thus more massive stars) per solar 
mass of molecular gas.  
A constant ratio of $L_{FIR}/M_{H_2}=100~L_\odot/M_\odot$
is an upper bound to what is
expected if massive star formation is primarily responsible
for the FIR luminosity \citep[see][]{sco91}, and
this plot can also offer a glimpse on whether the dust obscured 
powering source is massive stars or has to be a dust enshrouded AGN.
Among the 14 optically selected QSOs with measured molecular
gas contents (shown in filled circles),  only Mrk~1014 is found in
the area of high FIR luminosity and a large molecular gas
mass occupied by ULIRGs (empty circles).  The remaining
13 QSOs fall near the $L_{FIR}/M_{H_2}=10~L_\odot/M_\odot$
line which is characteristic of less luminous IR starbursts 
\citep[crosses; ][]{sanders91}.  This comparison suggests that
{\it optically selected QSOs are distinct from the
ULIRG population in general, even when their host galaxies are
fairly rich in molecular gas ($M_{H_2}=10^{9-10.5}M_\odot$)}.

The location of PDS~456 is distinct from most ULIRGs, 
less luminous IR starbursts, and even other optically selected QSOs
in Figure~\ref{fig:SFE}.  The uncertainty associated with the data 
points plotted in Figure~\ref{fig:SFE} is typically about 20\%, and 
the size of the error bars should be comparable or smaller than
the size of the symbols used.  Therefore the observed scatter in
this plot reflects substantial variations in the underlying physical 
processes (see below), far in excess of the measurement uncertainties.
PDS~456 has about 5 times more FIR luminosity than other
IR detected PG QSOs and LIRGs with comparable CO luminosity 
(molecular gas mass).  In fact, it is one of the objects 
showing the most extreme $L_{FIR}/M_{H_2}$ ($L_{FIR}/L'_{CO}$) ratio, and  
such a high $L_{FIR}/M_{H_2}$ ratio is reasonably expected if a large
fraction of the FIR luminosity arises from the ionizing
radiation originating from the AGN.  Another noteworthy object with 
a similarly extreme $L_{FIR}/M_{H_2}$ ratio is IRAS~08572+3915, which
is a well-known warm ULIRG ($S_{25\mu m}/S_{60\mu m}=0.23$ and 
$S_{60\mu m}/S_{100\mu m}= 1.64$) with a Seyfert/LINER optical spectrum
and shows IR-excess \citep[see][]{sanders88b,veilleux95,yun01}.  
From the near-IR and mid-IR imaging and spectroscopy, \citet{dudley97}, 
\citet{imanishi00}, and \citet{soifer00} have argued that dust-obscured
AGN activity is the dominant energy source for IRAS~08572+3915.
On the other hand, the failure to detect hard X-ray emission 
\citep[$L_{2-10kev}<4.4\times 10^{41}$ erg s$^{-1}$, ][]{risaliti00} 
or an unresolved VLBI radio source \citep{smith98} 
challenges the AGN interpretation.  Both X-ray and radio signatures
of an AGN can be obscured by a large column of neutral and ionized
gas, and we identify its proximity to the 
$L_{FIR}/M_{H_2}=100~L_\odot/M_\odot$ line in Figure~\ref{fig:SFE},
adjacent to the {\it bona fide} QSO PDS~456, as another evidence that AGN 
activity is the primary power source for IRAS~08572+3915.

The rare and unique nature of PDS~456 is further demonstrated by
the fact that the majority of the ``warm ULIRGs'', identified as possible
ULIRGs in transition to a QSO phase by \citet{sanders88b}, are
indistinguishable from other ULIRGs in Figure~\ref{fig:SFE}.
Nine out of the 12 ``warm ULIRGs'' identified by Sanders et al.
have been observed in CO thus far.  All nine have been detected
in CO, and they are already included as ULIRGs in 
Figure~\ref{fig:SFE}.  All but one of these sources cluster together
with the other ULIRGs in this figure.  IRAS~08572+3915 is the one 
exception, and indeed it stands out from the others with a
large $L_{FIR}/M_{H_2}$ ratio as noted already.
No CO measurements are available for the remaining
three (IRAS~01003$-$2238, IRAS~12071$-$0444, 3C~273), and we
cannot rule out the possibility that one or more of these three 
objects also have the large $L_{FIR}/M_{H_2}$ ratio similar to PDS~456. 
Regardless, we can safely conclude that objects like PDS~456 are rare,
even among the ``warm ULIRGs''.

Morphological studies of QSO host galaxies have found that not all
QSO hosts display evidence of a recent merger while 
IR-detected QSO hosts often do 
\citep[see][]{bahcall97,clements00,surace01}.  Because the size of the
accretion disk, mass accretion
rate, and total gas mass requirement for QSO activity are relatively
small, major mergers involving two massive gas-rich galaxies 
is not a necessary ingredient for the QSO phenomenon in general.  
However, given the high probability of finding more massive black 
holes in more massive progenitors
\citep{magorrian98,ferrarese00,gebhardt00} involved in major mergers
as well as the theoretical possibility of forming and/or growing
massive black holes within massive merger remnants 
\citep{weedman83,norman88,taniguchi99}, the likelihood of finding 
one or more massive black holes in an object undergoing a ULIRG 
phase should be naturally quite high.  
To this end, a recognizable trend is expected in 
Figure~\ref{fig:SFE} between ULIRGs and IR-detected QSOs.
Indeed ULIRGs (empty circles) and IR QSOs (filled circles) 
form a continuous distribution in Figure~\ref{fig:SFE} -- IR QSOs
follow a diagonal $L_{FIR}/M_{H_2}\sim 20~L_\odot/M_\odot$ line, which
connects smoothly to the nearly vertical distribution of ULIRGs 
near log $M_{H_2}\sim 10.5$.  If an IR-detected QSO represents
a later stage of evolution following the ULIRG phase, then 
this trend would indicate that the post-ULIRG
evolution occurs along a constant $L_{FIR}/M_{H_2}$ line
{\it characteristic of IR bright starbursts}.  

This observed trend between ULIRGs and IR QSOs is somewhat
surprising since it is not what is predicted by the existing 
ULIRG-QSO evolution scenarios. There is a plausible {\it a posteriori}
explanation to this trend, however.  Some scatter is 
seen in the distribution of the ULIRGs, LIRGs, and PG QSOs in
Figure~\ref{fig:SFE}, and one can interpret this as 
evidence for a wide range of ``efficiency'' in converting fuel
into luminosity.  Alternatively, one can interpret that this
wide range of observed ``efficiency'' reflects a range in the ratios
of dust processed luminosity contribution by AGN and starburst activity 
in these objects.  Among the optically visible QSOs, as
radiation pressure and winds clear out gas and dust from the immediate 
surroundings and reveal the optical AGN,  
the geometrical cross-section and the solid angle for 
dust heating by the AGN decrease rapidly \citep[see ][]{downes98},
and the AGN contribution to the FIR luminosity
diminishes quickly to the level where the underlying massive star 
formation again dominates the FIR luminosity.  In this scenario, a 
nearly vertical upturn in the distribution of ULIRGs near
$L_{CO}\sim 10^{10}$ K \kms\ pc$^2$ represents ubiquitous presence of
luminous AGNs among ULIRGs, providing a broad range of
additional luminosity.    Arp~220, which is thought
to be powered mostly by an intense starburst, thus appears
in the lower half of the ULIRG distribution while
Mrk~231 and Mrk~1014, both hosting luminous AGNs, appear near the 
top of the vertical ULIRG distribution where the elevated AGN 
contribution to the bolometric luminosity becomes significant.
Relatively little overlap between the two populations suggests
that the transition is quite rapid.

As noted earlier, PDS~456 and IRAS~08572+3915 are clearly displaced
from ULIRGs and IR QSOs in Figure~\ref{fig:SFE}, and they
represent an interesting challenge to the above ULIRG-QSO evolutionary
scenario.  One possible explanation is that
these objects follow a slightly different evolutionary path.  
Starting initially as a ULIRG with one or more luminous
AGNs (e.g. Mrk~231, Mrk~1014), one possible evolutionary path
they might follow is a rapid exhaustion of the gas reservoir via
a yet unknown mechanism, retaining only the compact, dense, and 
warm circum-AGN dust cocoon -- i.e., evolving nearly horizontally
to the left in Figure~\ref{fig:SFE}.  This evolutionary 
scenario is qualitatively similar to what was proposed by
\citet[][]{sanders88} although little detailed descriptions
were offered by these authors.  Few objects
are found along this possible evolutionary track, however, and this
is not likely a commonly followed evolutionary path.  

An alternative explanation is that objects like PDS~456 represent 
a brief transient phase of highly elevated FIR emission
in the ULIRG-QSO evolution.  One possible scenario is a momentary 
obscuration of the AGN by a episodic inflow of gas and dust, 
resulting in brief periods of dust coverage with a large solid angle
and brief increases in FIR luminosity by a factor of a few to ten. 
Such an event would materialize in Figure~\ref{fig:SFE} as a vertical 
upward displacement anywhere along the nominal evolutionary track. 
The rarity of objects like PDS~456 suggests that this period is 
brief compared with the duration of the IR QSO phase.
Numerical simulations of major mergers involving
two gas-rich disk galaxies \citep[e.g. ][]{barnes96,mihos96}
have shown that a large fraction of gas can be concentrated 
into the central 100 pc of the post-merger potential, fueling
luminous starbursts seen in the centers of ULIRGs.
These simulations also show that the gas inflow, which can last
over several hundred million years after the merger, can be lumpy, 
predicting a significant variation in the gas arrival rate over
the entire duration of merger that can last up to a billion years.
If a luminous AGN is already present, then this episodic nature
of the gas (and dust) inflow would predict a highly variable
obscuration of the central AGN -- 
the crossing time for a large gas/dust clump should be much
shorter ($<10^{6-7}$ years) than the galaxy merger time scale.  
In this scenario, objects like PDS~456 and IRAS~08572+3915 are
thus seen at a special moment in their ULIRG-QSO evolution,
rather than being a critical transition phase
as considered in \S~\ref{sec:intro}. Another important
consequence of this scenario is that it predicts 
a rare but distinct class of FIR sources with 
$L_{FIR}/M_{H_2}\sim 100~L_\odot/M_\odot$, spanning perhaps
the entire range of $L_{FIR}$ and $M_{H_2}$.  A small
number of ``IR-excess'' objects (characterized by a high
FIR-to-radio flux density ratio) identified by 
\citet[][]{yun01} among the IRAS detected galaxies may indeed represent 
this rare population.  As mentioned already,
IRAS~08572+3915 is such an object.  Another of the few well studied
``IR-excess'' objects, NGC~4418, has a high SFE of $L_{FIR}/M_{H_2}=69$
$L_\odot/M_\odot$ \citep[log $M_{H_2}=9.17$,
log $L_{FIR}=11.01$, ][]{sanders91} and is indeed
thought to host a dust-enshrouded AGN \citep[see ][]{spoon01}. 

The unusually high $L_{FIR}/M_{H2}$ ratio observed for objects
like PDS~456 and IRAS~08572+3915 may have a different explanation,
unrelated to the ULIRG-QSO evolution scenario.  
Whatever the process, it has to account for both the large IR 
luminosity and the large dust/gas content simultaneously.
The paucity of objects with the extreme $L_{FIR}/M_{H2}$ ratio of 
PDS~456 suggests that it also 
has to be a rare process, and its duration must be quite short
compared with the length of the luminous phase for QSOs.  

\section{Summary}

We report a new, detailed multi-wavelength observational study 
of PDS~456, the most luminous QSO in the local universe ($z\lesssim 0.3$).
The $0.6''$ resolution \kband\ image obtained using the Keck telescope
shows three compact $m_K\sim 16.5$ ($M_K\sim -21$) sources located
$\sim 10$ kpc southwest of the QSO, but little detail on the host 
galaxy is revealed.  The near-IR color of these objects are
consistent with their being at the same redshift as PDS~456.
These imaging results suggest that the host galaxy of PDS~456 
may be interacting or merging with one or more companions.  

Observations using
the OVRO millimeter array has revealed a narrow CO (1--0) line 
($\sigma = 77$ \kms; FWHM = 181 \kms) centered at $z=0.1849$.
The narrow line width is consistent with a face-on geometry
for the circum-AGN gas disk as also suggested by the optically
visible QSO.  A total H$_2$ mass of $9\times 10^9 M_\odot$ is inferred
for the host galaxy from the CO luminosity. 
The most accurate astrometry of the QSO is derived
with subarcsecond accuracy using the 
new high resolution VLA continuum image at 1.2 GHz, and new
radio continuum and millimeter continuum measurements
are used to constrain the long wavelength SED of PDS~456.
The eight times larger radio continuum flux density 
than expected from the radio-FIR correlation is a 
clear indication that a radio AGN is present in PDS~456.

Our 850 $\mu$m photometric observations of PDS~456 using the
SCUBA camera on JCMT did not produce a detection at the
level of 7.5 mJy ($3\sigma$).  On the other hand, this upper 
limit is sufficient to rule out an Arp220-like host with
copious amount of cold dust for PDS~456.  
The total amount of cold dust present in the host system does not 
exceed about one half of the dust present in Arp~220.

The SEDs of PDS~456, I~Zw~1, and 3C~273 are remarkably similar 
between optical and FIR wavelengths, including the pronounced
mid-IR enhancement which is a well known characteristic of AGNs
\citep{deG85,yun01}.  At the same time, the
SED between FIR and radio wavelengths is similar to that of an
ultraluminous galaxy Arp~220, suggesting a dual
QSO-like and ULIRG-like nature.  The observed FIR, X-ray, and CO 
properties of PDS~456 suggest that the AGN activity dominates its 
luminosity.  

While a face-on geometry for the dust and gas disks in PDS~456 
and Mrk~231 seems quite secure, a surprising new result is that
a broad inference of a face-on geometry for nuclear gas disks 
among QSO host systems is {\it not} supported by
the observed distribution CO line widths.  Among the
small number of ULIRGs and PG QSOs examined,  
QSO hosts show comparable CO line widths as ULIRGs,
and the possibility of optical AGNs
viewed through gaps in a large scale gas/dust ring or disk
at an arbitrary inclination angle cannot be ruled out.  We conclude that 
a face-on geometry for gas and dust distribution around optically 
visible QSOs, as expected from the AGN ``unification'' scheme,
may be an over-simplification.

\begin{figure}
\plotone{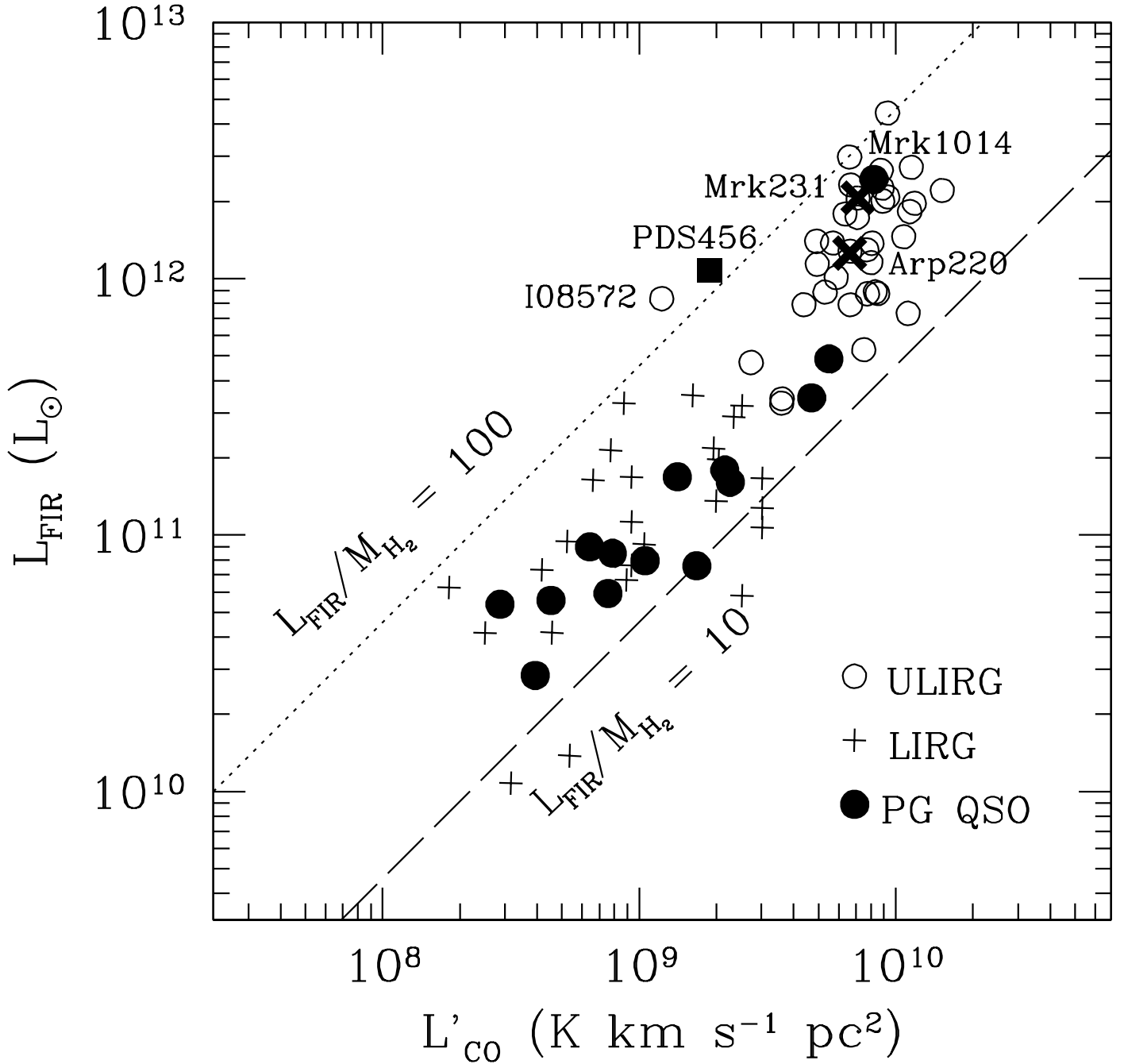}
\figcaption{FIR luminosity versus $L'_{CO}$.  PDS~456 is shown
as a filled square while 14 optically selected PG QSOs
with available CO measurements \citep[][]{alloin92,evans01,scoville03} 
are shown in filled circles.  Typical uncertainty in all of these
measurements are about 20\%, and the size of the errorbars are
comparable or smaller than the size of the symbols shown.
The prototype ULIRG Arp~220 and
IR QSO/ULIRG Mrk~231 are identified using a thick X.  
Luminous infrared galaxies \citep[crosses;][]{sanders91}
and ULIRGs \citep[empty circles;][]{solomon97} 
are also plotted for comparison.  $L_{FIR}$ is derived
using the definition given in Table~1 of \citet{sanders96}.
$L'_{CO}$ can be converted to H$_2$ mass
($M_{H_2}$) using the standard CO-to-H$_2$ conversion
given by \citet{young91}.  High temperature and density in 
intense starburst and circum-AGN regions may cause an over-estimate 
of the gas mass by a factor of a few \citep[see][]{sco97,downes98},
but this effect should apply uniformly to all objects compared here.
\label{fig:SFE}}
\end{figure}

PDS~456 and IRAS~08572+3915
are two objects with an extremely large $L_{FIR}/M_{H_2}$
ratio, $\sim100~L_\odot/M_\odot$, which is the upper
bound of what can be reasonably expected from a dust enshrouded 
starburst. A comparison of the gas mass-to-light ratio
clearly shows that optically selected IR QSOs are distinct from the
ULIRG population in general, even when their host galaxies are
fairly rich in molecular gas ($M_{H_2}=10^{9.0-10.5}M_\odot$).
Relatively little overlap between the two populations suggests
that the transition is quite rapid.
A continuous track formed by ULIRGs and IR QSOs
in the ``star formation efficiency'' plot (Figure~\ref{fig:SFE})
suggests a transition in the luminosity contribution by dust obscured
AGN from a starburst, in the context of the ULIRG-QSO evolution 
scenario.  We further infer that the FIR luminosity of the IR QSOs is 
mostly dominated by the underlying starburst activity under the
ULIRG-QSO evolution scenario.   
We also propose that objects like PDS~456 represent a 
special transient phase in the ULIRG-QSO evolution scenario where the
central AGN is momentarily obscured by an episodic inflow of
gas and dust.  As a consequence, we predict a rare but distinct 
population of dust obscured AGNs with 
$L_{FIR}/M_{H_2}\gtrsim 100~L_\odot/M_\odot$ covering an 
entire range of FIR luminosity.

In summary, PDS~456, the most luminous QSO in the local
universe, does not possess the FIR and gas properties consistent
with the ULIRG-QSO evolution scenario proposed by \citet{sanders88b}
despite its ULIRG-like FIR and CO luminosity and classical QSO 
signatures of broad emission lines and large X-ray luminosity.
The continuous distribution of ULIRGs and IR QSOs in the
``star formation efficiency'' plot (Figure~\ref{fig:SFE}) is
broadly consistent with a ULIRG-QSO evolution scenario,
and the FIR emission associated with the IR QSOs are dominated
by the underlying starburst under this scenario.  PDS~456 clearly
deviates from the observed ULIRG-QSO trend in this plot.  The paucity of
PDS456-like objects in the local universe strongly suggests that
PDS~456 is a rare, exceptional object, and the duration 
of the physical process governing its properties must be short
compared with the length of the luminous QSO phase.

\acknowledgments

The authors are grateful to J. Lowenthal for his assistance with the
initial examination of the Keck \kband\ imaging data.  
The authors also thank the Keck Observatory staff who made the 
$K$-band observations possible.
This publication makes use of data products from the Two Micron All
Sky Survey, which is a joint project of the University of Massachusetts
and the Infrared Processing and Analysis Center/California Institute
of Technology, funded by the National Aeronautics and Space
Administration and the National Science Foundation.
This research has made use of the NASA/IPAC Extragalactic Database 
(NED) which is operated by the Jet Propulsion Laboratory, California Institute
of Technology, under contract with the National Aeronautics and 
Space Administration.  Support for this work was provided in part by a grant 
from the K. T. and E. L. Norris Foundation and NSF grant AST 99-81546.
N. Reddy was supported by the National Science Foundation Graduate
Research Fellowship.  The JCMT is operated by the Joint Astronomy Centre, 
on behalf of the UK Particle Physics and Astronomy Research Council, 
the Netherlands Organisation for Scientific Research, and the National 
Research Council of Canada.



\clearpage 

\clearpage 

\clearpage 

\clearpage 

\clearpage 

\clearpage 

\clearpage 

\clearpage 

\begin{deluxetable}{lccccl}
\tablecaption{Summary of the $K$-band Sources in the PDS~456 Field 
\label{tab:K}}
\tablewidth{0pt}
\tablehead{
\colhead{} &
\colhead{Object} &
\colhead{$\alpha$} &
\colhead{$\delta$} &
\colhead{$m_K$} \\
\colhead{} &
\colhead{} &
\colhead{(J2000)} &
\colhead{(J2000)} &
\colhead{}}
\startdata
& QSO & 17:28:19.78 & $-$14:15:56.12 & 9.89$\pm$ 0.02  \\
& K1  & 17:28:19.64 & $-$14:15:57.54 & 16.3$\pm$ 0.2  \\
& K2+K3  & 17:28:19.69 & $-$14:15:59.02 & 15.6$\pm$ 0.1  \\
\enddata
\end{deluxetable}


\begin{deluxetable}{lccccc}
\tablecaption{Spectral Energy Distribution Data for PDS~456 
\label{tab:sed}}
\tablewidth{0pt}
\tablehead{
\colhead{} &
\colhead{$\lambda$} &
\colhead{$\nu$} &
\colhead{$S_\nu$} &
\colhead{Reference} \\
\colhead{} &
\colhead{} &
\colhead{(GHz)} &
\colhead{(mJy)} &
\colhead{}}
\startdata
& {\bf 25 cm} & {\bf 1.2} & {\bf 24$\pm$5}      & {\bf this work}  \\
& 21 cm & 1.4 & 30.4          & 1  \\
& 21 cm & 1.4 & $22.7\pm 3.3$ & 2  \\
& 6 cm  & 4.85 & 8.23         & 1  \\
& {\bf 3.6 cm} & {\bf 8.5} & {\bf 4.6$\pm$0.9}  & {\bf this work}  \\
& {\bf 3000 $\mu$m} & {\bf 100} & {\bf $<$0.6}       & {\bf this work}  \\
& {\bf 850 $\mu$m} & {\bf 350} & {\bf $<$7.5}       & {\bf this work}  \\
& 100 $\mu$m & 3000 & $420\pm140$  & 1  \\
& 100 $\mu$m & 3000 & $<882$       & 3  \\
& 80 $\mu$m & 3750  & $510\pm110$  & 1  \\
& 60 $\mu$m & 5000  & $930\pm69$   & 3  \\
& 50 $\mu$m & 6000  & $610\pm110$  & 1  \\
& 25 $\mu$m & 12000 & $750\pm41$   & 3  \\
& 25 $\mu$m & 12000 & $550\pm120$  & 1  \\
& 12 $\mu$m  & 25000 & $514\pm24$   & 3  \\
& 12 $\mu$m  & 25000 & $280\pm30$   & 1  \\
& 4.8 $\mu$m & 62500 & $225\pm36$   & 4  \\
& 3.8 $\mu$m & 78900 & $180\pm23$   & 4  \\
& 2.2 $\mu$m & 136400 & $78\pm6$    & 4  \\
& {\bf 2.2 $\mu$m} & {\bf 136400} & {\bf 73$\pm$7}    & {\bf this work}  \\
& 2.159 $\mu$m & 139000 & $78.0\pm0.8$    & 5  \\
& 1.662 $\mu$m & 178300 & $38.6\pm0.6$   & 5  \\
& 1.65 $\mu$m & 181800 & $39\pm3$   & 4  \\
& 1.25 $\mu$m & 240000 & $26\pm2$   & 4  \\
& 1.235 $\mu$m & 242800 & $24.8\pm0.3$   & 5  \\
\enddata
\tablerefs{
(1) Reeves et al. 2000; (2) Condon et al. 1998; (3) IRAS ADDSCAN;
(4) Simpson et al. 1999; (5) 2MASS
}

\end{deluxetable}


\begin{thebibliography}{}

\bibitem[Alloin et al.(1992)]{alloin92} Alloin, D., Barvainis, R., 
	Gordon, M. A., Antonucci, R. R. J. 1992, A\&A, 265, 429

\bibitem[Andreani, Franceschini, \& Granato(1999)]{andreani99}
	Andreani, P., Franceschini, A., Granato, G. 1999, MNRAS, 306, 161

\bibitem[Bahcall et al.(1997)]{bahcall97} Bahcall, J. N., Kirhakos, S.,
Saxe, D. H., Schneider, D. P. 1997, ApJ, 479, 642


\bibitem[Barnes \& Hernquist(1996)]{barnes96} Barnes, J. E., \&
Hernquist, L. 1996, ApJ, 471, 115


\bibitem[Blain et al.(1999)]{blain99} Blain, A. W., Jameson, A., 
	Smail, I., Longair, M. S., Kneib, J.-P., Ivison, R. 1999
	MNRAS, 309, 715

\bibitem[Bryant \& Scoville(1996)]{bryant96} Bryant, P. M., Scoville,
N. Z. 1996, ApJ, 457, 678 

\bibitem[Burstein \& Heiles(1982)]{burstein82} Burstein, D., \& Heiles,
	C. 1982, AJ, 87, 1165

\bibitem[Carilli et al.(2001)]{carilli01} Carilli, C. L., Bertoldi,
F., Omont, A., Cox, P., McMahon, R. G., \& Isaak, K. 2001, AJ, 122, 1679

\bibitem[Clements(2000)]{clements00} Clements, D. L. 2000, MNRAS,
	311, 833
 
\bibitem[Condon et al.(1998)]{Con98} Condon, J. J., Cotton, W. D., Greison, 
	E. W., Yin, Q. F., Perley, R. A., et al. 1998, AJ, 115, 1693

\bibitem[Courvoisier(1998)]{courvoisier98} Courvoisier, T. J.-L. 1998,
	A\&ARv, 9 , 1

\bibitem[de Grijp et al.(1985)]{deG85} de Grijp, M. H. K., Miley, G. K., 
	Lub, J., de Jong, T.  1985, Nature, 314, 240

\bibitem[Djorgovski et al.(1995)]{djorg95} Djorgovski, S., Soifer, B. T.,
	Pahre, M. A., Larkin, J. E., Smith, J. D., et al. 1995, ApJ, 438, L13

\bibitem[Downes \& Solomon(1998)]{downes98} Downes, D., \& Solomon, P. M.
	1998, ApJ, 507, 615

\bibitem[Dudley \& Wynn-Williams(1997)]{dudley97} Dudley, C. C., \&
	Wynn-Williams, C. G. 1997, ApJ, 488, 720

\bibitem[Dunlop et al.(2003)]{dunlop03} Dunlop, J. S., McLure, R. J.,
	Kukula, M. J., Baum, S. A., O'Dea, C. P., Hughes, D. H. 2003,
	MNRAS, 340, 1095

\bibitem[Evans et al.(2001)]{evans01} Evans, A. S., Frayer, D. T.,
	Surace, J. A., Sanders, D. B. 2001, AJ, 121, 1893

\bibitem[Ferrarese \& Merritt(2000)]{ferrarese00} Ferrarese, L., \&
	Merritt, D. 2000, ApJ, 539, L9

\bibitem[Gebhardt et al.(2000)]{gebhardt00} Gebhardt, K., Kormendy, J.,
	Ho, L. C., Bender, R., Bower, G. et al.
	2000, ApJ, 539, L13

\bibitem[Genzel et al.(1998)]{genzel98} Genzel, R., Lutz, D., Sturm, E.,
	Egami, E., Kunze, D. et al. 1998, ApJ, 498, 579

\bibitem[Haas et al.(1998)]{haas98} Haas, M., Chini, R., Meisenheimer,
	K., Stickel, M., Lemke, D. et al. 1998, ApJ, 503, L109

\bibitem[Haas et al.(2000)]{Haas00} Haas, M., M\"{u}ller, S. A. H., Chini,
	R., Meisenheimer, K., Klaas, U. et al. 2000, A\&A, 354, 453

\bibitem[Hughes et al.(1993)]{hughes93} Hughes, D. H., Robson, E. I.,
	Dunlop, J. S., Gear, W. K. 1993, MNRAS, 263, 607

\bibitem[Hughes et al.(1998)]{hughes98} Hughes, D. H., Serjeant, S.,
	Dunlop, J., Rowan-Robinson, M., Blain, A. et al. 1998, 
	Nature, 394, 241

\bibitem[Imanishi \& Dudley(2000)]{imanishi00} Imanishi, M., \&
	Dudley, C. C. 2000, ApJ, 545, 701

\bibitem[Levenson, Weaver, \& Heckman(2001)]{levenson01} Levenson, N. A., 
	Weaver, K. A., Heckman, T. M. 2001, ApJ, 550, 230

\bibitem[Madau et al.(1996)]{madau96} Madau, P., Ferguson, H. C., 
Dickinson, M. E., Giavalisco, M., Steidel, C. C., Fruchter, A. 1996,
MNRAS, 283, 1388

\bibitem[Magorrian et al.(1998)]{magorrian98} Magorrian, J., Tremaine,
S., Richstone, D., Bender, R., Bower, G., et al. 1998, AJ, 115, 2285


\bibitem[Martel et al.(2003)]{martel03} Martel, A. R., Ford, H. C.,
	Tran, H. D., Illingworth, G. D. Krist, J. E. et al. 2003,
	AJ, 125, 2964

\bibitem[Mas-Hesse et al.(1995)]{mas-hesse95} Mas-Hesse, J. M., 
	Rodr\'{i}guez-Pascual, P. M., Fern\'{a}ndez de C\'{o}rdova,
	Mirabel, I. F., Wamsteker, W., Makino, F., Otani, C. 1995,
	A\&A, 298, 22

\bibitem[Matthews \& Soifer(1994)]{matthews94} Matthews, K., Soifer, B. T.
1994, {\it Experimental Astronomy}, 3, 77

\bibitem[McLeod \& McLeod(2001)]{mcleod01} McLeod, K. K., McLeod, B. A.
	2001, ApJ, 546, 782

\bibitem[McLure et al.(1999)]{mclure99} McLure, R. J., Kukula, M. J.,
	Dunlop, J. S., Baum, S. A., O'Dea, C. P., Hughes, D. H. 1999,
	MNRAS, 308, 377
 
\bibitem[Mihos \& Hernquist(1996)]{mihos96} Mihos, J. C. \& Hernquist, L. 
1996, ApJ, 464, 641

\bibitem[Norman \& Scoville(1988)]{norman88} Norman, C., \& Scoville,
N. 1988, ApJ, 332, 124

\bibitem[Omont et al.(2001)]{omont01} Omont, A., Cox, P., Bertoldi, F.,
McMahon, R. G., Carilli, C., \& Isaak, K. 2001, A\&A, 374, 371

\bibitem[Ptak et al.(2003)]{ptak03} Ptak, A., Heckman, T., Strickland,
	D., Levenson, N. A., Weaver, K. 2003, ApJ, 592, 782

\bibitem[Reeves et al.(2000)]{reeves00} Reeves, J. N., O'Brien, P. T.,
Vaughan, S., Law-Green, D., Ward, M. et al. 2000, MNRAS, 312, L17


\bibitem[Risaliti et al.(2000)]{risaliti00} Risaliti, G., Gilli, R.,
	Maiolino, R., \& Salvati, M. 2000, A\&A, 357, 13

\bibitem[Rowan-Robinson(1995)]{rowan95} Rowan-Robinson, M. 1995,
	MNRAS, 272, 737

\bibitem[Sakamoto et al.(1999)]{sakamoto99} Sakamoto, K., Scoville,
	N. Z., Yun, M. S., Corsas, M., Genzel, R., \& Tacconi, L. J.
	1999, ApJ, 514, 68

\bibitem[Sanders et al.(1988a)]{sanders88} Sanders, D. B., Soifer, B. T.,
	Elias, J. H., Madore, B. F., Matthews, K. et al. 1988a
	ApJ, 325, 74

\bibitem[Sanders et al.(1988b)]{sanders88b} Sanders, D. B., Soifer, B. T.,
	Elias, J. H., Neugebauer, G., \& Matthews, K. 1988b
	ApJ, 328, L35

\bibitem[Sanders et al.(1989)]{sanders89} Sanders, D. B., Phinney, E. S.,
Neugebauer, G., Soifer, B. T., Matthews, K.  1989, ApJ, 347, 29

\bibitem[Sanders et al.(1991)]{sanders91} Sanders, D. B., Scoville,
	N. Z., Soifer, B. T. 1991, ApJ, 370, 158

\bibitem[Sanders \& Mirabel(1996)]{sanders96} Sanders, D. B., \&
	Mirabel, I. F. 1996, ARAA, 34, 749

\bibitem[Scoville \& Soifer(1991)]{sco91} Scoville, N. Z., \& Soifer, 
	B. T. 1991,
	in {\it Massive Stars in Starbursts,} eds. C. Leitherer, N.R.
	Walborn, T.M. Heckman, \& C.A. Norman (Cambridge Univ. Press: New
	York), p 233.

\bibitem[Scoville et al.(1991)]{scoville91} Scoville, N. Z., Sargent, A. I.,
	Sanders, D. B., \& Soifer, B. T. 1991, ApJ, 366, L5

\bibitem[Scoville et al.(1992)]{Sco92} Scoville, N. Z., Carlstrom,
  J. C., Chandler, C. J., Phillips, J. A., Scott, S. L., Tilanus,
  R. P., \& Wang, Z. 1992, PASP, 105, 1482

\bibitem[Scoville, Yun, \& Bryant(1997)]{sco97} Scoville, N. Z.,
Yun, M. S., \& Bryant, P. M. 1997, ApJ, 484, 702

\bibitem[Scoville et al.(2003)]{scoville03} Scoville, N. Z.,
Frayer, D. T., Schinnerer, E., Christopher, M. 2003, ApJ, 585, L105

\bibitem[Shepherd et al.(1994)]{Shepherd94} Shepherd, M.C., Pearson, 
  T.J., \& Taylor, G.B. 1994, BAAS, 26, 987

\bibitem[Simpson et al.(1999)]{simpson99} Simpson, C., Ward, M.,
O'Brien, P., Reeves, J. 1999, MNRAS, 303, L23 

\bibitem[Shaver et al.(1996)]{shaver96} Shaver, P. A., Wall, J. V.,
	Kellermann, K. I., Jackson, C. A., Hawkins, M. R. S. 1996,
	Nature, 384, 439

\bibitem[Skrutskie et al.(1997)]{skrutskie97} Skrutskie, M. F.,
	et al. 1997, Proc. Workshop ``The Impact of Large Scale
	Near-IR Sky Surveys'', ed. Garz\'{o}n F. et al. (Dordrecht:
	Kluwer), 25

\bibitem[Smith et al.(1998)]{smith98} Smith, H. E., Lonsdale, C. J.,
	\& Lonsdale, C. J. 1998, ApJ, 492, 137

\bibitem[Soifer et al.(2000)]{soifer00} Soifer, B. T., Neugebauer, G.,
	Matthews, K., Egami, E., Becklin, E. E. et al. 2000, AJ, 119, 509

\bibitem[Solomon et al.(1997)]{solomon97} Solomon, P. M., Downes, D.,
Radford, S. J. E., Barrett, J. W. 1997, ApJ, 478, 144

\bibitem[Spoon et al.(2001)]{spoon01} Spoon, H. W. W., Keane, J. V.,
	Tielens, A. G. G. M., Lutz, D., Moorwood, A. F. M. 2001,
	A\&A, 365, L353

\bibitem[Surace, Sanders, \& Evans(2001)]{surace01} Surace, J. A.,
	Sanders, D. B., \& Evans, A. S. 2001, AJ, 122, 2791
 
\bibitem[Tacconi et al.(2002)]{tacconi02} Tacconi, L. J., Genzel,
	R., Lutz, D., Rigopoulou, D., Baker, A. J., et al. 2002,
	ApJ, 580, 73

\bibitem[Taniguchi, Ikeuchi, \& Shioya(1999)]{taniguchi99} 
Taniguchi, Y., Ikeuchi, S., \& Shioya, Y. 1999, ApJ, 514, L9

\bibitem[Torres et al.(1997)]{torres97} Torres, C. A. O., Quast, G. R.,
Coziol, R., Jablonski, F., de La Reza, R. et al. 1997, ApJ, 488, L19

\bibitem[Veilleux et al.(1995)]{veilleux95} Veilleux, S., Kim, D.-C.,
	Sanders, D. B., Mazzerella, J. M., \& Soifer, B. T. 1995,
	ApJS, 98, 171

\bibitem[Vignali et al.(2000)]{vignali00} Vignali, C., Comastri, A.,
Nicastro, F., Matt, G., Fiore, F., Palumbo, G. G. C. 2000, A\&A, 362, 69

\bibitem[Weedman(1983)]{weedman83} Weedman, D. W. 1983, ApJ, 266, 479

\bibitem[Young \& Scoville(1991)]{young91} Young, J. S., \& Scoville,
N. Z. 1991, ARAA, 29, 581

\bibitem[Yun \& Scoville(1998)]{yun98} Yun, M. S., \& Scoville, N. Z.
1998, ApJ, 507, 774

\bibitem[Yun et al.(2001)]{yun01} Yun, M. S., Reddy, N. A., Condon, J. J.
2001, ApJ, 554, 803

\bibitem[Yun \& Carilli(2002)]{yun02} Yun, M. S., \& Carilli, C. L. 2002,
ApJ, 568, 88

\end{thebibliography}
\end{document}